\documentclass[aps,prd,floatfix,amsfonts,amssymb,amsmath,twocolumn,
   notitlepage,nofootinbib,groupedaddress,superscriptaddress]{revtex4}
\usepackage{graphicx}
\usepackage{slashed}
\newcommand{\diff}{\text{\rm d}}\newcommand{\euler}{\text{\rm e}}
\newcommand{\imu}{\text{\rm i}}

\usepackage{times}

\begin{document}
\title{Dense baryonic matter: \\constraints from recent neutron star observations}
\author{Thomas Hell}
\affiliation{Physik-Department, Technische Universit\"{a}t M\"{u}nchen, D-85747 Garching, Germany}
\affiliation{ECT*, Villa Tambosi, I-38123 Villazzano (Trento), Italy}
\author{Wolfram Weise}
\affiliation{Physik-Department, Technische Universit\"{a}t M\"{u}nchen, D-85747 Garching, Germany}
\affiliation{ECT*, Villa Tambosi, I-38123 Villazzano (Trento), Italy}

\date{\today}

\begin{abstract}

Updated constraints from neutron star masses and radii impose stronger restrictions on the equation of state for baryonic matter at high densities and low temperatures. The existence of two-solar-mass neutron stars rules out many soft equations of state with prominent ``exotic" compositions. The present work reviews the conditions required for the pressure as a function of baryon density in order to satisfy these constraints. Several scenarios for sufficiently stiff equations of state are evaluated. The common starting point is a realistic description of both nuclear and neutron matter based on a chiral effective field theory approach to the nuclear many-body problem. Possible forms of hybrid matter featuring a quark core in the center of the star are discussed using a three-flavor Polyakov--Nambu--Jona-Lasinio (PNJL) model. It is found that a conventional equation of state based on nuclear chiral dynamics meets the astrophysical constraints. Hybrid matter generally turns out to be too soft unless additional strongly repulsive correlations, e.g. through vector current interactions between quarks, are introduced. The extent to which strangeness can accumulate in the equation of state is also discussed.

\end{abstract}

\maketitle

\begin{section}{Introduction}

The investigation of compressed baryonic matter is one of the persistently important themes in the physics of strongly interacting many-body systems. While high-energy heavy-ion collisions probe the transition from 
the hadronic phase to deconfined quark-gluon matter at high temperatures and relatively low baryon chemical potentials,
the access to ``cold" and dense baryonic matter comes primarily through observations of neutron stars in which central core densities several times the density of normal nuclear matter can be reached. 

Two remarkable examples of massive neutron stars have recently emerged. One of those is the radio pulsar J1614--2230 with a mass $M = (1.97\pm0.04) M_\odot$ \cite{Demorest:2010bx}. Even heavier neutron stars were occasionally discussed in the literature (e.\,g., \cite{Lattimer:2010uk} and references therein), but this one is special because of the high accuracy of its mass determination made possible by the particular edge-on configuration (an inclination angle of almost $90^\circ$) of the binary system consisting of the pulsar and a white dwarf. Given this configuration, a pronounced Shapiro-delay signal of the neutron star's pulses could be detected. In the meantime a second neutron star has been found with a comparable, accurately determined mass (J0348+0432 with $M = (2.01\pm0.04) M_\odot$) \cite{Antoniadis:2013pzd}, further strengthening the case.

The established existence of two-solar-mass neutron stars rules out many equations of state (EoS) that are too soft to stabilize such stars against gravitational collapse. On the other hand, some selected equations of state based entirely on conventional nuclear degrees of freedom are able to develop a sufficiently high pressure so that the condition to reach $2 M_\odot$ can be satisfied  \cite{AP:1997, APR:1998, Engvik:1996}.

The present work performs an updated analysis of the constraints on the EoS of strongly interacting baryonic matter provided by these observations.  Traditionally, the primary source of information is the mass-radius relation of the star calculated using the Tolman-Oppenheimer-Volkov equations \cite{PhysRev.55.374,1934PNAS...20..169T,PhysRev.55.364} with a given EoS as input. The empirical restrictions on neutron star radii are less severe than those on the mass. Nonetheless, the quest for a stiff EoS at high baryon densities persists as a common theme throughout this investigation. Earlier related studies that include less stringent constraints from heavy-ion collisions in addition to those from neutron star properties are summarized in Ref.\,\cite{klaehn:2006}. In the present work we do not discuss heavy-ion collisions.

An essential condition to be fulfilled is the following: the known properties of normal nuclear matter must be considered as a prerequisite for the construction of any realistic EoS, together with the requirement of consistency with advanced many-body computations of pure neutron matter (see e.g. \cite{GCRSW:2014, RMP:2014}). This latter important constraint has so far not been respected by equations of state routinely used in supernova simulations \cite{LS:1991,STOS:1998,SHT:2011,KTHS:2013}. 

Neutron star matter interpolates between the extremes of isospin-symmetric nuclear matter and pure neutron matter. The fraction of protons added to the neutron sea is controlled by beta equilibrium. The passage from $N = Z$ matter to neutron-rich matter as it emerges in the core of the star is driven by detailed properties of the isospin-dependent part of the nuclear interaction. These isospin-dependent forces also determine the evolution of the nuclear liquid-gas phase transition from isospin-symmetric matter towards the disappearance of this phase transition around $Z/N \simeq 0.05$. Such properties of the phase diagram of highly asymmetric nuclear matter provide further guidance and constraints that we incorporate in our analysis.

At the interface between low-energy quantum chromodynamics (QCD) and nuclear physics, chiral effective field theory (ChEFT) has become the framework for a successful description of the nucleon-nucleon interaction and three-body forces, as well as for the nuclear many-body problem (see Refs.~\cite{Ep:2006, Epelbaum:2009, Machleidt:2011, HKW:2013} for recent reviews). 
ChEFT is our starting point for a systematic approach to nuclear and neutron matter at densities (and temperatures) well within the hadronic sector of QCD, the one governed by confinement and spontaneous chiral symmetry breaking. The ChEFT approach is used here to set the boundary values, at normal nuclear densities, for the construction of the EoS at higher densities. As will be demonstrated, a sufficiently stiff EoS supporting a two-solar-mass neutron star does indeed result from in-medium ChEFT with ``conventional" (nucleon and pion) degrees of freedom plus three-body forces. Options for a transition to quark matter at very high baryon densities will be examined using a three-flavor Polyakov--Nambu--Jona-Lasinio (PNJL) model. It turns out to be unlikely, however, that such a quark component, even if existent in the deep interior of the star, will be of observable significance. Furthermore, the possible role of hyperons will briefly be  discussed, again with the condition in mind that their admixture should not soften the EoS so much that it falls short of supporting a two-solar-mass neutron star.  

The aim of the present paper is then twofold: first, to establish boundaries and constraints that any equation of state should fulfill in view of the recent astrophysical observations; secondly, to construct a realistic EoS with a firm foundation in the (chiral) symmetry breaking pattern of low-energy QCD. In Section\,\ref{massradiusconstraintssection} the mass constraint together with (less restrictive) constraints on neutron-star radii are summarized in order to impose general limitations for the EoS of neutron star matter. In this context the neutron star crust is briefly discussed.
In Section\,\ref{eossection} the equations of state for symmetric and asymmetric nuclear matter and for pure neutron matter are constructed within the framework of in-medium ChEFT. This includes the resummation of short-range interaction ladders to all orders in the large neutron-neutron scattering length. Comparisons with state-of-the-art many-body calculations of neutron matter will be displayed. Section\,\ref{neutronstarsection} is then devoted to astrophysical implications of these EoS results. A summary and conclusions are presented in Section\,\ref{conclusionssection}. 

\end{section}

\begin{section}{Empirical constraints from neutron stars}\label{massradiusconstraintssection} 

Apart from the mass measurements discussed in the introduction, this section briefly reviews and summarizes empirical constraints on neutron star radii and their implications. Thereafter it is shown how the two-solar-mass pulsars (J1614--2230 and J0348-0432), in combination with the (considerably less accurate) radius restrictions, define conditions for acceptable equations of state for neutron star matter.

\begin{subsection}{Neutron star radii}\label{nsradiussection}

In this work we consider constraints on neutron star radii from {\bf several} independent sources. The first one, Refs.~\cite{Steiner:2010fz,Steiner:2012xt,Lattimer:2013hma}, following earlier studies in Refs.~\cite{Ozil:2009,Ozil:2010}, is based on a statistical analysis of the mass-radius curves of four X-ray bursters (EXO~1745--248, 4U~1608--522, 4U~1820--30, KS~1731--260), and four quiescent low-mass X-ray binaries (neutron stars in the  globular clusters 47~Tuc, $\omega$~Cen, M13 and NGC~6397). Reference~\cite{Lattimer:2013hma} amends the previous analyses by considering in addition the low-mass X-ray binaries in the globular clusters NGC~6304 and M28. Analyzing the X-ray spectra of the neutron stars and assuming that all objects have hydrogen atmospheres, one arrives at typical radii, $R(1.4)$, for $1.4$-solar-mass neutron stars ranging from $10.4$ to $12.9\,{\rm km}$ (95\,\% confidence level) \cite{Steiner:2012xt} and $11.4$ to $12.8\,{\rm km}$ (90\,\% confidence level) \cite{Lattimer:2013hma}. A recently updated analysis gives $R(1.4) = 12.1 \pm 1.1$ km. According to Ref.~\cite{Lattimer:2013hma} radii of neutron stars having masses between $0.8\,M_\odot$ and $2.0\,M_\odot$ all lie in a band between $10.9$ and $12.7\,{\rm km}$, and a similar band ranging from $11.2$ to $12.8$ km is quoted in \cite{LS:2014} for individual stars with masses between $1.2\,M_\odot$ and $1.8\,M_\odot$. An analysis performed in Ref.~\cite{Guillot:2013wu} considering the same objects as in Ref.~\cite{Lattimer:2013hma}, but assuming a  constant radius for all neutron stars, leads to $R = 9.1^{+1.3}_{-1.5}\,{\rm km}$. However, the statistical method used in that analysis results in a radius range that is smaller than the accepted radii assigned to most of the individual neutron stars under consideration. 

As a second source we refer to the neutron star radius constraints provided by Fig.~6 of Ref.~\cite{Truemper2013}. This detailed analysis features four independently determined curves of constraints that, taken together, form a rhombic area in the mass-radius plot. In combination with the two-solar-mass condition a triangular area remains, bounded by radii $11.5 \lesssim R \lesssim 14.5$ km. Within the given uncertainties, all acceptable equations of state should generate mass-radius trajectories that pass through this triangle. These radius constraints are deduced from the following specific cases: the light-curve oscillations of the X-ray burster XTE J1814--338 \cite{Bhattacharyya:2004pp}; the thermal spectrum of the radio-quiet isolated neutron star RXJ~1856--3754 as discussed in Ref.~\cite{Thoma:2003xq} (recalling, however, the analysis of Ref.~\cite{Ho:2007} that arrives at a smaller radius than \cite{Thoma:2003xq}); the 90\,\%-confidence analysis using a hydrogen-atmosphere model to fit the spectra of neutron stars in the globular cluster 47~Tuc \cite{Rybicki:2005id,Bogdanov:2006ap} (with the added comment in \cite{Truemper2013} that this deduced radius may be a lower limit); and finally, the mass-shedding limit calculated from the spinning period of the fastest known pulsar, J1748--2446ad.

Significant uncertainties associated with all of those deduced neutron star radii are of course 
to be kept in mind. In the following the two sources of information and analysis just mentioned will be used in parallel. The resulting constraints cover altogether broad bands of radii for which we can assume that they represent a reasonably conservative estimate of uncertainties. 

\end{subsection}

\begin{subsection}{Mass-radius relation}

Given an equation of state (EoS) relating pressure and energy density,
the  mass-radius curves for neutron stars are determined by solving the Tolman-Oppenheimer-Volkoff (TOV) equation. This equation describes the structure of a spherically symmetric star composed of isotropic material with corrections from general relativity \cite{PhysRev.55.374,1934PNAS...20..169T,PhysRev.55.364}:
\begin{equation}\label{tov}
	\begin{aligned}
	\dfrac{\diff P(r)}{\diff r}&=-\dfrac{\mathcal{G}}{r^2 c^2}\left[\epsilon(r)+P(r)\right]\left[M(r)+4\pi r^3\dfrac{P(r)}{c^2}\right]\\
	&\quad\times\left[1-\dfrac{2\mathcal{G} M(r)}{c^2 r}\right]^{-1}\,.
	\end{aligned}
	\end{equation}
Here $\mathcal{G}$ is the gravitational constant, $c$ denotes the speed of light\footnote{In all subsequent sections units with $c = 1$ will be used.}, $r$ is the radial coordinate,  and $\epsilon(r)$ and $P(r)$ are the energy density and pressure, respectively. Moreover, $M(r)$ is the total mass inside a sphere of radius $r$. It is related to the energy density by
	\begin{equation}\label{massdgl}
		\dfrac{\diff M(r)}{\diff r}=4\pi r^2\dfrac{\epsilon(r)}{c^2}\,.
	\end{equation}
Eqs.~\eqref{tov} and \eqref{massdgl} supplemented by an EoS, $P=P(\epsilon)$, determine completely the structure of a static (non-rotating), spherical neutron star. The commonly chosen initial boundary conditions for the integration of the TOV equation are the energy density in the core of the neutron star, $\epsilon(0)=\epsilon_c$, and $M(0)=0$. The radius, $R$, of the neutron star is given by the condition $\epsilon(R)=\epsilon_\text{Fe}$, where the energy density on the surface of the star has dropped down to that of atomic iron, $\epsilon_\text{Fe}=7.9\,\text{g}/\text{cm}^{3}=4.4\cdot10^{-12}\,\text{MeV}/\text{fm}^{3}$. The neutron star mass is
	\begin{equation}\label{nstarmass}
		M\equiv M(R)={4\pi\over c^2}\int_0^R \diff r\, r^2 \epsilon(r)\,\, ,
	\end{equation}
the total mass measured by the gravitational field felt by a distant observer. 
\end{subsection}

\begin{subsection}{Neutron star equation of state: \\ 
~~~~~constraints from observables}
\label{nsconstraintstotaleossection}

The primary purpose of this preparatory subsection is to provide minimally model-dependent constraints on the equation of state for neutron star matter, in a similar way as previously described in Refs. \cite{Steiner:2010fz, Steiner:2012xt, Lattimer:2013hma, LS:2014, Ozil:2009, Ozil:2010, Hebeler:2010jx, Hebeler:2013nza}. A detailed modeling of the EoS, satisfying these constraints and extrapolating to neutron star core densities, will then be presented in the subsequent section guided by in-medium ChEFT as a basic framework, with extensions to possible hybrid matter at the highest densities.

Solving the TOV equation requires the knowledge of the EoS in the entire neutron star, including the low-density crust region at its surface. The outer crust is associated with densities $\varrho\lesssim\varrho_d$ below the neutron-drip point, $\varrho_d\approx 10^{-3}\,\varrho_0$ in units of nuclear saturation density, $\varrho_0=0.16\,{\rm fm}^{-3}$. The structure of this outer crust region is quite well established \cite{Baym:1971pw}. The inner crust is less well understood \cite{Negele:1971vb}. In the transition region to a uniform nuclear medium (in the density range $0.2\,\varrho_0\lesssim\varrho\lesssim0.5\,\varrho_0$) extended clusters of so-called ``pasta" phases \cite{Ravenhall:1983uh,Watanabe:2011yz} might be formed.

In order to describe this multifacet structure of the neutron star's crust (not covered by our explicit calculations) we use the empirical equation of state as given in Ref.~\cite{Haensel:2004nu} for the low-density region.  This EoS is fitted to a Skyrme-Lyon EoS \cite{Douchin:2001sv} and to experimental data for neutron-rich nuclei according to Refs.~\cite{Baym:1971pw,Haensel:1993zw}. In the following, we refer to this crust EoS as ``SLy". 

At a density of about $0.5\,\varrho_0$ the nuclei dissolve and turn into a uniform medium of neutrons with a small admixture of protons in the outer core region of the neutron star. In order to interpolate between regions from lower densities up to around $\rho_0$, we adopt the ChEFT-based EoS determined in Refs.~\cite{Fiorilla:2011sr, HKW:2013} (FKW), assuming at this point {for simplicity} a (constant) proton fraction of 10\,\%. (The detailed evaluation of the proton fraction via beta equilibrium is performed in Section \ref{neutronstarsection}). The FKW EoS is matched to the SLy EoS at their intersection point, $\epsilon_0\approx 118\,{\rm MeV}/{\rm fm}^3$ corresponding to a density $\varrho \approx 0.75\, \varrho_0$.
\begin{figure}[htbp]
\begin{center}
 \includegraphics[width=.45\textwidth]{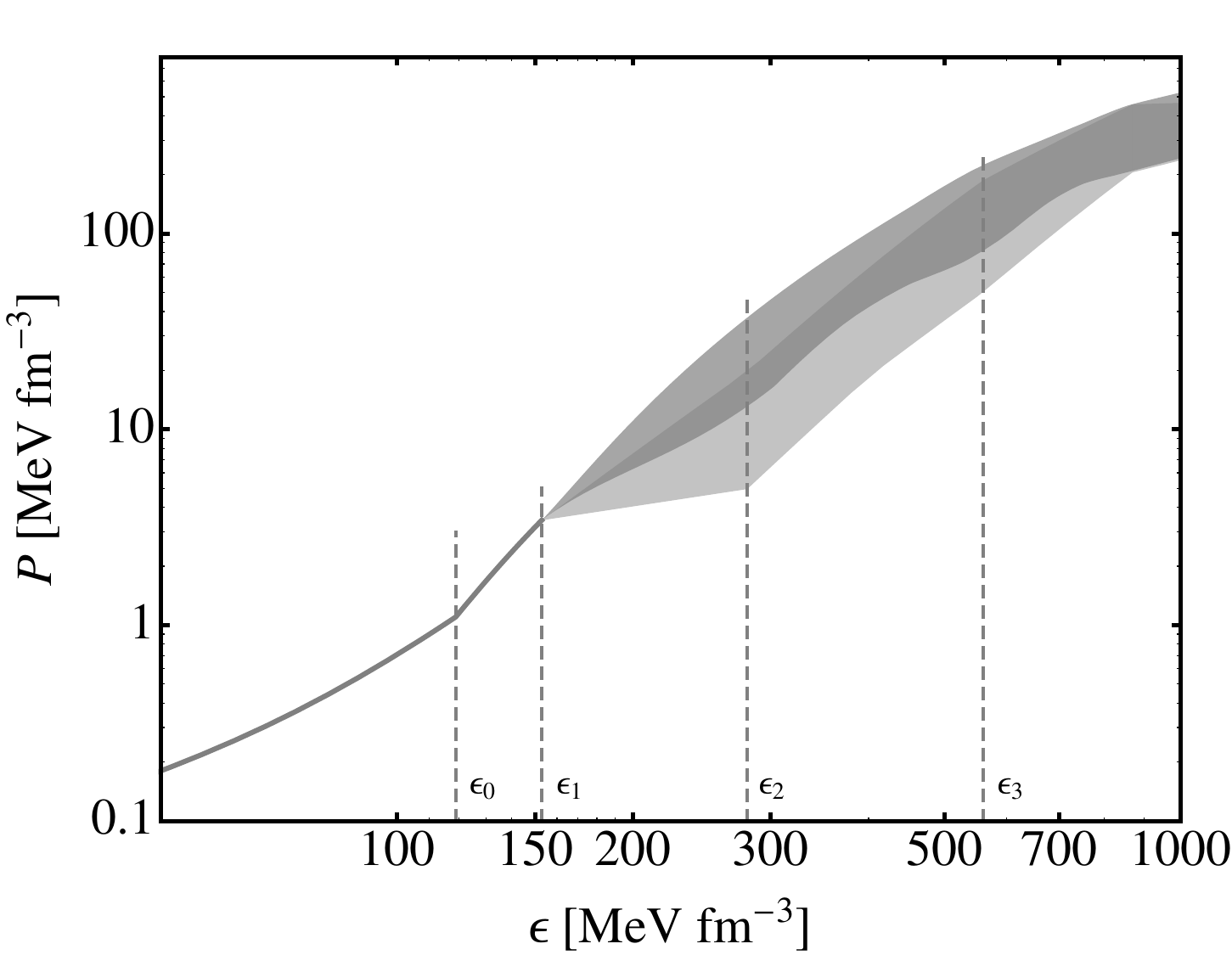}
\end{center}
\vspace{-.5cm}
\caption{Allowed regions for the equation of state $P(\epsilon)$ as dictated by neutron star observables. The   upper (dark grey) area takes into account the limitations as given by Tr\"umper \cite{Truemper2013} and constraints from causality. The lower (light grey) band uses, in addition to the two-solar-masses constraint, a permitted radius window $11.0$--$12.5\,{\rm km}$ from \cite{Steiner:2010fz,Steiner:2012xt,Lattimer:2013hma}. For energy densities smaller than $\epsilon_1$ and $\epsilon_0$ the FKW and SLy EoS, respectively, are used. The matching points $\epsilon_1,\epsilon_2,\epsilon_3$ of the polytropes in Eq. (\ref{eq:polytropes}) are also shown in the figure. 
}
\label{constraintseos}
\end{figure}

The extrapolation to the high-density domain of the equation of state is parametrized using three polytropes fitted sequentially to one another (in a way similar to the procedure pursued in Refs.~\cite{Hebeler:2010jx,Hebeler:2013nza}): $P=K_i\,\varrho^{\varGamma_i}$, $i\in\{1,2,3\}$. The equation of state for each of the branches is
	\begin{equation}
			\epsilon=a_i\left(\dfrac{P}{K_i}\right)^{1/\varGamma_i}+\dfrac{1}{\varGamma_i-1}\,P~~~~~ (i = 1,2,3)\,,
\label{eq:polytropes}
	\end{equation}
where the $a_i$ are constants determined by the continuity of $\epsilon=\epsilon(P)$.  It turns out that three polytropes are sufficient \cite{Read:2008iy} in order to represent a large variety of models for dense nuclear matter. We use the FKW EoS up to an energy density $\epsilon_1=153\,{\rm MeV}/{\rm fm}^3$ corresponding to nuclear saturation density. The polytropes are then introduced in the ranges between $\epsilon_1$ and $\epsilon_2=280\,{\rm MeV}/{\rm fm}^3$, $\epsilon_2$ to $\epsilon_3=560\,{\rm MeV}/{\rm fm}^3$ and at energy densities larger than $\epsilon_3$. The parameters $\varGamma_i$ and $K_i$ are fixed such that the equation of state is continuous at the matching points. Instead of varying $\varGamma_1$ we vary the pressure $P_2=P(\epsilon_2)$. Following Ref.~\cite{Read:2008iy} the parameters $P_2$, $\varGamma_2$ and $\varGamma_3$ are varied in the following ranges:
 		\begin{equation}\label{polytroprange}
			\begin{aligned}
		\log_{10}~{P_2\,\,{\rm fm}\over{\rm MeV}}&=0.7+n_1\cdot0.1\le1.6\,,\\
		\varGamma_2&=1.2+n_2\cdot 0.65\le 3.8\,,\\
		\varGamma_3&=1.3+n_3\cdot0.8\le 3.7\,,
		\end{aligned}
		\end{equation}
with $n_1,n_2,n_3\in\mathbb{N}$. The constraints from neutron star masses and radii then translate into a limited band area of $P(\epsilon)$. Any acceptable EoS must lie within this belt.

Combining the SLy EoS for $\epsilon<\epsilon_0$, the FKW EoS for $\epsilon_0\le\epsilon<\epsilon_1$ and the three polytropic equations of state for $\epsilon_1\le\epsilon<\epsilon_2$, $\epsilon_2\le\epsilon<\epsilon_3$, and $\epsilon\ge\epsilon_3$, the TOV equation is solved for each set \eqref{polytroprange}. We accept a parameter set $(P_2,\varGamma_2,\varGamma_3)$ if the resulting mass-radius curve reaches or passes beyond the two-solar-mass limit dictated by J1614--2230 and J0348+0432, and if it is within the range of radii suggested by Steiner, Lattimer, Brown \cite{Steiner:2010fz,Steiner:2012xt,Lattimer:2013hma} or, alternatively, passes through the constraining triangle as given by Tr\"umper \cite{Truemper2013}. For the Steiner-Lattimer-Brown constraints we keep all parameter sets that generate mass-radius curves exceeding the two-solar-mass limit in the radius range $11.0$--$12.5\,{\rm km}$ \emph{and} crossing the $M=1.4\,M_\odot$ line in the radius window $10.5$--$13.0\,{\rm km}$ \cite{Steiner:2010fz,Steiner:2012xt,Lattimer:2013hma}. We ensure that causality is not violated, i.\,e. the speed of sound, $v_{\rm s}$, satisfies the condition
		\begin{equation}
			v_{\rm s}=\sqrt{\dfrac{\diff P}{\diff \epsilon}}\le 1\,.
		\end{equation}
The result of this analysis is presented in Fig.~\ref{constraintseos}. The bands comprise all polytropes that meet the constraints dictated by the neutron star observables and causality. These emerging ``allowed" corridors are consistent with the results reported in Ref.~\cite{Hebeler:2013nza}. 

It is of interest to point out that state-of-the-art EoS's computed using advanced quantum Monte Carlo methods \cite{GCRSW:2014}, as well as the time-honored EoS resulting from a variational many-body calculation \cite{APR:1998} (APR),
both  pass the test of being within the allowed $P(\epsilon)$ region, once three-nucleon forces are included and the nuclear symmetry energy is constrained around $E_{sym} \simeq 33$ MeV. Notably, these equations of state work with ``conventional" (baryon and meson) degrees of freedom.

\end{subsection}

\end{section}

\begin{section}{Equations of state}\label{eossection}

This section deals with the construction of an EoS for baryonic matter at densities relevant to the description of the neutron star core. The framework is chiral effective field theory (ChEFT), the approach based on the spontaneously broken chiral symmetry of low-energy QCD. ChEFT has been applied successfully to the nuclear many-body problem and its thermodynamics, for symmetric nuclear matter, pure neutron
matter and varying proton fractions Z/A between these extremes (see Ref.\,\cite{HKW:2013} for a recent review and references therein). At high baryon densities, the possible appearance of hybrid matter with admixtures of deconfined quark degrees of freedom will also be explored using a Nambu and Jona-Lasinio model including strange quarks. It will be demonstrated, however, that a significant quark matter component is not likely to appear even in the very central region of the neutron star core, given the new observational constraints requiring a sufficiently stiff equation of state.

\begin{subsection}{Chiral effective field theory}\label{chefteos}

In-medium ChEFT incorporates the essentials of low-energy pion-nucleon and pion-pion interactions together with the Pauli principle and a systematically structured hierarchy of nucleon-nucleon forces that include one- and two-pion exchange dynamics plus important three-body correlations. In the present work we use an equation of state for neutron star matter (neutron matter with an admixture of protons) based on three-loop in-medium ChEFT calculations of nuclear and neutron matter \cite{HKW:2013,Fritsch:2005,Fiorilla:2011sr}. 

The starting point is the chiral meson-baryon effective Lagrangian in its isospin SU(2) sector, with pions as the ``light" (Goldstone boson) degrees of freedom coupled to nucleons as ``heavy" sources. 
This Lagrangian is organized as an expansion in powers of pion momentum (derivatives of the pion field) and pion mass (the measure of explicit chiral symmetry breaking by the small non-zero $u$- and $d$-quark masses):
\begin{equation}
{\cal L}_{\pi N} = {\cal L}_{\pi N}^{(1)} + {\cal L}_{\pi N}^{(2)} + \dots ~~.
\end{equation}
At leading order we have
\begin{equation}
{\cal L}_{\pi N}^{(1)} =  \bar{\Psi}\Big[i\gamma_{\mu}(\partial^{\mu} +\Gamma^{\mu})- M_0 + 
g_A \gamma_{\mu}\gamma_5\, u^{\mu}\Big]\Psi \,,\label{eq:LeffN}
\end{equation}
with the isospin doublet Dirac field of the nucleon, $\Psi = (u, d)^\top$. The vector and axial vector quantities 
\begin{eqnarray}
\Gamma^{\mu} & = & {1\over 2}[\xi^{\dagger},\partial^{\mu}\xi] = {i\over 4f_\pi^2}
 \, \vec \tau \cdot (\vec \pi \times \partial^{\mu}\vec \pi) + ...~~, \\
u^{\mu} & = & {i\over 2}\{\xi^{\dagger},\partial^{\mu}\xi\}= - {1\over 2f_\pi}\, \vec \tau
\cdot \partial^{\mu}\vec \pi + ...~~,
\end{eqnarray}
involve the isovector pion field $\vec{\pi}$ via $\xi = \exp[({\rm i}/2f_\pi)\vec{\tau}\cdot\vec{\pi}]$. 
The last steps in the preceding equations result when expanding $\Gamma^{\mu}$ and $u^{\mu}$ to 
leading order in the pion field. Up to this point the only parameters that enter are the 
nucleon mass $M_0$, the nucleon axial vector coupling constant $g_A$ and the pion decay 
constant $f_\pi$, all to be taken at first in the chiral limit. The pion decay constant 
plays the role of an order parameter for spontaneous chiral symmetry breaking. It sets a characteristic
scale, $4\pi f_\pi \sim 1$ GeV. The effective field theory is designed to work at excitation energies and momenta small compared to that scale.

At next-to-leading order, ${\cal L}_{\pi N}^{(2)}$, the chiral symmetry breaking 
quark mass term enters. It has the effect of shifting the nucleon mass from
its value in the chiral limit to the physical mass. The nucleon sigma term
\begin{equation}
\sigma_N = m_q\frac{\partial M_N}{\partial m_q} =
\langle N | m_q(\bar{u}u + \bar{d}d) |N\rangle
\end{equation}
measures the contribution of the non-vanishing quark mass, $m_q =
\frac{1}{2}(m_u + m_d)$, to the nucleon mass $M_N$. Its empirical value is in the range
$\sigma_N \simeq (45 \pm 8)$ MeV and has been deduced \cite{GLS91} by
extrapolation of low-energy pion-nucleon data using dispersion relation techniques.
Up to this point, the $\pi N$ effective Lagrangian, expanded to second order
in the pion field, has the form
\begin{eqnarray}
{\cal L}_{\text{eff}}^{N} & = & \bar{\Psi}(i\gamma_{\mu}\partial^{\mu} - M_N)\Psi - 
{g_A \over 2f_{\pi}} \bar{\Psi}\gamma_{\mu}\gamma_5\vec\tau\,\Psi \cdot\partial^{\mu}
\vec \pi  \nonumber \\ &-& {1 \over 4f_{\pi}^2} \bar{\Psi}\gamma_{\mu} \vec \tau\,\Psi
\cdot (\vec \pi\times \partial^{\mu} \vec \pi\,)\nonumber \\
 &+&{\sigma_N\over  2f_\pi^2}\,\bar{\Psi}
\Psi\,\vec \pi^{\,2} + \dots~~, 
\label{effLagrangian}
\end{eqnarray}
where we have not shown a series of additional terms involving $(\partial^{\mu} \vec\pi)^2$ 
that appear in the complete Lagrangian ${\cal L}_{\pi N}^{(2)}$.
These terms come with low-energy constants $c_{3,4}$ encoding physics at smaller 
distances or higher energies. These constants need to be fitted to experimental data, 
e.g. from pion-nucleon scattering. 

The ``effectiveness" of such an effective field theory relies on the proper 
identification of the active low-energy degrees of freedom. Pion-nucleon scattering is 
known to be dominated by the $p$-wave $\Delta(1232)$ resonance with spin and isospin 
3/2. The excitation energy of this resonance, given by the mass difference 
$\Delta=M_\Delta - M_N \simeq 293\,$MeV is small, just slightly larger than twice the pion mass.
If the physics of the $\Delta(1232)$ is absorbed in low-energy 
constants such as $c_{3,4}$ of an effective theory that works with pions and nucleons only, 
the limit of applicability of such a theory is 
narrowed down to an energy-momentum range small compared to $\Delta$. 
The effective Lagrangian is therefore often extended \cite{Fritsch:2005,HHK97,pasca1,pasca2} by incorporating the 
$\Delta(1232)$ isobar as an explicit degree of freedom, and this is the version of ChEFT that
we use here to construct an EoS for neutron star matter. 

The pion-nucleon vertices entering Eq.\,(\ref{effLagrangian}) generate a systematically organized hierarchy
of pion exchange mechanisms in the nucleon-nucleon interaction: one-pion exchange at leading order (LO), 
two-pion exchange processes at next-to-leading order (NLO) and so forth \cite{Ep:2006, Epelbaum:2009, Machleidt:2011}. These explicitly calculated long- and intermediate-range parts are supplemented by 
NN contact terms that encode short distance dynamics not resolved in detail at small momenta far below
the chiral symmetry breaking scale, $4\pi f_\pi$, of order 1 GeV. The constants associated with these contact terms are parameters to be fixed and fine-tuned by comparison with empirical data. In the standard version of ChEFT, terms involving important p-wave pion-nucleon scattering information through the low-energy constants $c_{3,4}$ appear at next-to-next-to-leading order (N$^2$LO). Three-body NNN forces also emerge for the first time at  N$^2$LO. As mentioned, the version we use in this work is the one with $\Delta(1232)$ degrees of freedom treated explicitly. In this case, two-pion exchange processes involving intermediate $\Delta$ excitations are promoted from N$^2$LO to NLO, rescaling the constants $c_{3,4}$ and improving
the convergence of the approach. The importance of the $N\rightarrow\Delta$ transition in generating the very large spin-isospin polarizability of the nucleon is underlined in this way. This also emphasizes the significance of virtual $\Delta$ excitations in providing a prominent part of the central attraction in the $2\pi$ exchange NN force at intermediate distances, as well as an important piece of the three-body interaction.

This scheme has been applied successfully to the description of symmetric and asymmetric nuclear matter as well as pure neutron matter \cite{HKW:2013,Fritsch:2005,Fiorilla:2011sr}. In particular, nuclear thermodynamics, the liquid-gas phase transition, its evolution as a function of the proton fraction $Z/A$ and its disappearence in neutron matter, are well reproduced. The isospin dependence of explicit two-pion exchange processes in the nuclear medium plays an important role in this context. In-medium ChEFT provides a systematic way to handle such mechanisms, including the action of the Pauli principle in the presence of filled Fermi seas of neutrons and protons with varying proportions. 
The Pauli principle is implemented through the in-medium nucleon propagator,
\begin{eqnarray}
G(E,\vec{p}\,) = {i\over E-{\vec{p}\,^2\over 2M_N} + i\epsilon} 
- 2\pi \delta\left(E - {\vec{p}\,^2\over 2M_N}\right)\Theta(p)
\end{eqnarray} 
where
\begin{eqnarray}
\Theta(p) = {1+\tau_3\over 2}\,\theta(k_F^p - |\vec{p}\,|) + 
{1-\tau_3\over 2}\,\theta(k_F^n - |\vec{p}\,|)~,
\end{eqnarray} 
and $k_F^{p,n}$ are the proton and neutron Fermi momenta, respectively. Intermediate and long-range pion exchange dynamics (Fock terms from one-pion exchange and all explicit two-pion exchange processes in the presence of the medium) are computed up to three-loop order in the energy density. Contact terms (subject to resummations as described in \cite{kaiser:2011}) are adjusted to properties of symmetric nuclear matter
(the empirical binding energy per nucleon and the equilibrium density) and to the symmetry energy at $k_F^0 = 1.36$ fm$^{-1}$.

The ``small" parameters, in addition to pion mass and momentum, now include the Fermi momenta, $k_F^{p,n}/ 4\pi f_\pi\ll 1$. The energy density is derived as an expansion in powers of Fermi momenta and generally written as: 
\begin{eqnarray}
\epsilon(k_F^p, k_F^n) = \epsilon_0(k_F) + \delta^2 A_2(k_F) + \dots~~,
\end{eqnarray} 
introducing the asymmetry parameter $\delta = (\varrho_n - \varrho_p)/\varrho$ with the neutron and 
proton densities,
\begin{eqnarray}
\varrho_{n,p} = {(k_F^{n,p})^3\over 3\pi^2}~,  
\end{eqnarray} 
and the total baryon density, $\varrho = \varrho_p + \varrho_n$. For symmetric nuclear matter,
$ \varrho = 2k_F^3/(3\pi^2)$.
Symmetric nuclear matter and pure neutron matter correspond to the limiting cases $\delta = 0$ 
and $\delta = 1$, respectively. A good approximation for $\delta \lesssim 1$ relevant 
for neutron star matter, with a small admixture of protons controlled by beta equilibrium, is given
by extrapolating around the neutron matter limit, $\delta = 1$, using the $\delta^2$ term.

\begin{figure}[htbp]
\begin{center} 
 \includegraphics[width=.53\textwidth]{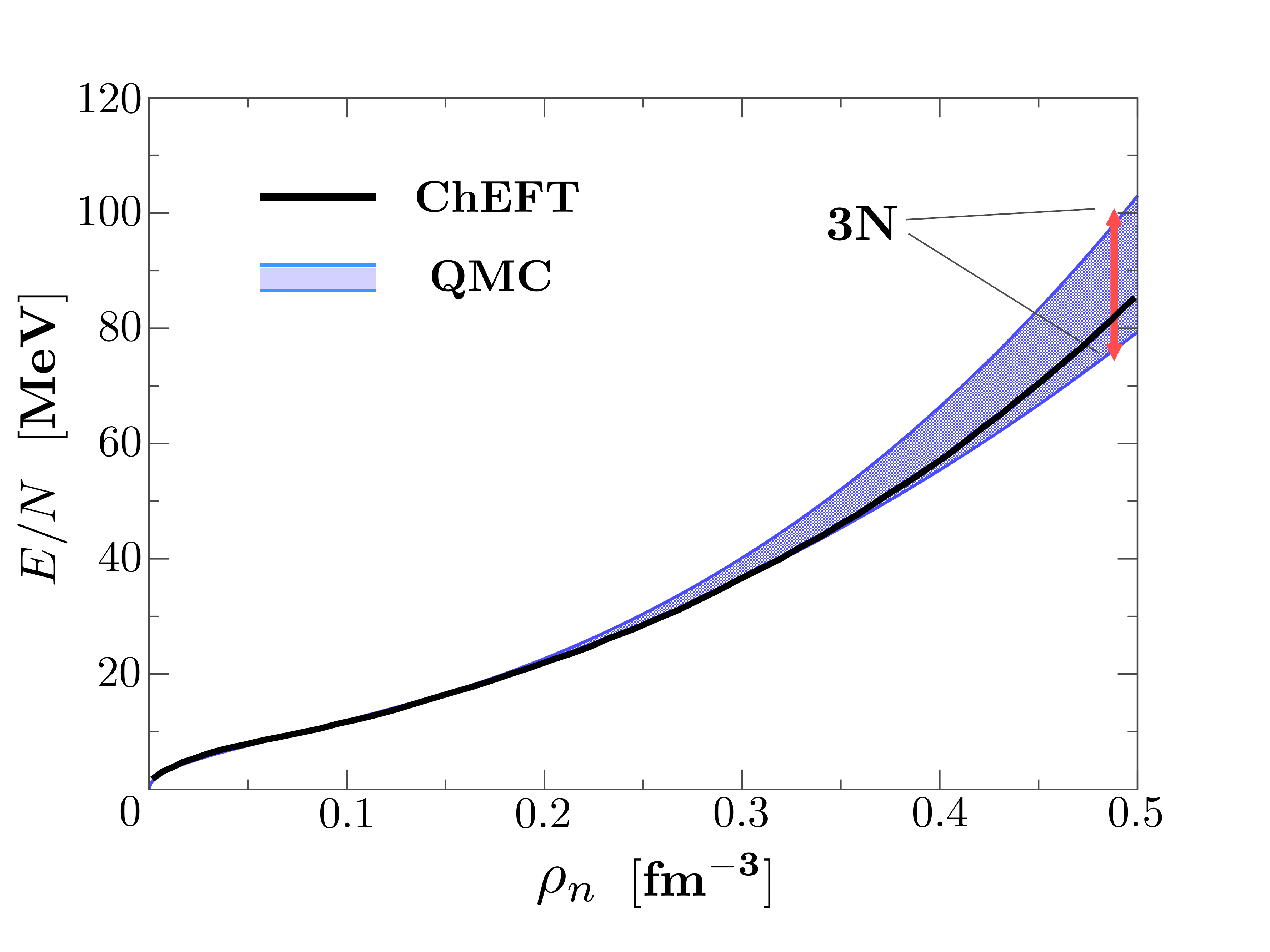}
\end{center}
\vspace{-.5cm}
\caption{Energy per particle, $E/N = \epsilon/\rho_n - M_n$, for pure neutron matter as a function of density $\rho_n$. Solid curve: ChEFT result \cite{HKW:2013b,Fritsch:2005,Fiorilla:2011sr} used in the present work. Blue shaded area: results deduced from Quantum Monte Carlo (QMC) computations reviewed in \cite{GCRSW:2014}, using different models of the three-neutron force (3N).
}
\label{nmatter}
\end{figure}
The ChEFT equation of state used in this work operates with a limited set of altogether four parameters associated with contact terms (and derivatives thereof) representing short-distance dynamics, plus a cutoff, $\Lambda = 0.75$ GeV, in dispersion integral representations of two-pion exchange loop diagrams.\footnote{This cutoff is not to be confused with the momentum cutoff usually associated with chiral low-momentum interactions, $\Lambda_{\text{low-k}}\sim 2$ fm$^{-1}$.} Two of those parameters, denoted $B_3$ and $B_5$, appear in conjunction with contact terms contributing at order $k_F^3/M_N^2$ and $k_F^5/M_N^4$ to the energy per particle in symmetric nuclear matter. The two remaining ones, $B_{3n}$ and $B_{5n}$, are specific to neutron matter.\footnote{Including resummations of contact terms, the optimal input values from best fits to equilibrium nuclear matter and to the symmetry energy are: $B_3=-1.36, B_5 = -17.7, B_{3n} = 0, B_{5n} = -2.2$. This set is used in the present work.}

The EoS derived from in-medium chiral EFT can be tested by comparing the
result for pure neutron matter with sophisticated and advanced many-body calculations. 
Figure \ref{nmatter} shows such a comparison with an EoS based on recent Quantum Monte Carlo (QMC) computations reviewed in Ref.\,\cite{GCRSW:2014}. The QMC equation of state includes three-body interactions which play an important role in the extrapolation to high densities. Uncertainties associated with these three-neutron forces, shown the figure, are discussed in detail in Ref.\,\cite{GCRSW:2014}. Within these uncertainties the quality of the agreement between the ChEFT and QMC equations of state is good  
even at densities as high as three times the density of normal nuclear matter and beyond. At $\rho_n \sim 3\,\rho_0$ the neutron Fermi momentum, $k_F^n \sim 2.4$ fm$^{-1}$, continues to be appreciably smaller than the chiral symmetry breaking scale of order $4\pi f_\pi \sim$ 1 GeV, rendering the ChEFT expansion in powers of $x = k_F/4\pi f_\pi$ still meaningful\footnote{Note that $x\sim 0.5$ even at densities as high as $\rho_n \sim 5\,\rho_0$.}. The sensitivity to convergence issues in the chiral expansion of the energy per particle starts at order $x^4$ and involves even higher powers of $x$. The only exception to this scheme
is the case of reducible two-nucleon processes such as iterated one-pion exchange (dominated by the in-medium second-order tensor force), for which the relative scaling factor is $M_N\,k_F/(4\pi f_\pi)^2$. Such diagrams are calculated exactly up to three-loop order in the energy density. 

Elaborating further on questions of convergence, it is instructive to compare the (perturbative) ChEFT expansion in the nuclear medium with calculations that start from a chiral meson-nucleon Lagrangian based on a linear sigma model plus short-distance interactions, combined with a (non-perturbative) functional renormalization group (FRG) approach \cite{Drews:2013,Drews:2014}. The latter takes into account leading subclasses of in-medium pionic fluctuations and nucleonic particle-hole excitations to all orders.
The close similarity of those ChEFT and FRG results, both for symmetric nuclear matter \cite{Drews:2013} and for neutron matter \cite{Drews:2014} holds up to at least three times the density of nuclear matter.

Uncertainties related to the previously mentioned cutoff in the ChEFT approach have been examined by
varying this cutoff in the range $0.6$ GeV $\le \Lambda \le 0.9$ GeV, i.e. by $\pm 20\%$ around the standard value, $\Lambda = 0.75$ GeV.  The resulting changes in $E/N$ are marginal at $\varrho_0 = 0.16\,\,\text{fm}^{-3}$, about $10\%$ at $\varrho_n = 3\,\varrho_0$ and $15\%$ at $\varrho_n = 5\,\varrho_0$.
 
\end{subsection}

\begin{subsection}{Quark matter: PNJL model with vector interaction}

At very high baryon densities the principal possibility exists that nucleons dissolve into a sea of quarks. In this subsection, quark matter is described using the Polyakov-loop-extended Nambu and Jona-Lasinio (PNJL) model with $N_{\rm f}=2+1$ quark flavors, taking into account two degenerate light (up and down) quarks with masses $m_u=m_d$, and a heavier strange quark with mass $m_s$. The PNJL approach has been developed and discussed extensively in the literature \cite{Fukushima:2003fm,fukushima-2004-591,Ratti:2007jf,Roessner:2006xn,Rossner:2007ik,Hell:2008cc,
Fukushima:2008wg}. 

Neutron stars are ``cold" systems, with temperatures $T$ typically below a few MeV. Given the $u$ and $d$ current-quark masses of the same order, it is useful to prepare the EoS of quark matter at finite $T$ and then take the limit $T \rightarrow 0$ (done here also in view of neutron star cooling issues that are, however, not part of the present work). 

The starting point is the (Euclidean) action of the (local) PNJL model: 
\begin{equation}\label{spnjl}
\begin{aligned}
		\mathcal{S}_\text{PNJL}&=\int_{0}^\beta\diff \tau\int\diff^3 x\,\bar q(x)\left(-\imu\gamma_\nu D^\nu+\gamma_0\,\hat\mu+\hat m\right)q(x)\\&\quad+\int_{0}^\beta\diff \tau\int\diff^3 x\,\mathcal{L}_\text{int}+\beta V\,\mathcal{U}(\Phi[A],\bar\Phi[A];T)\,.
\end{aligned}
\end{equation}
where $q(x)=(u(x),d(x),s(x))^\top$ is the three-flavor quark field and $\hat m={\rm diag}_f(m_u,m_d,m_s)$ denotes the (current) quark mass matrix. We work in the isospin limit with $m_u=m_d$. Quark chemical potentials are incorporated in the matrix $\hat\mu={\rm diag}_f(\mu_u,\mu_d,\mu_s)$. 

The interaction part of the Lagrangian, $\mathcal{L}_{\text{int}}$, is given as:
	\begin{equation}\label{3flint}
	\begin{aligned}
	\mathcal{L}_\text{int}&=\dfrac{1}{2}\, G\sum_{a=0}^8\left[(\bar q\,\lambda^a q)^2+(\bar q\,\imu\gamma_5\lambda^a q)^2\right]+\mathcal{L}_v\\&\quad-K\left[\det\left(\bar q(1+\gamma_5) q\right)+\det\left(\bar q(1-\gamma_5) q\right)\right]\,.
	\end{aligned}
	\end{equation}
The first term in the first line describes the chirally invariant combination of scalar and pseudoscalar  interactions between quarks, with coupling strength $G$ of dimension (length)$^2$. The flavor SU(3) Gell-Mann matrices $\lambda_i ~(i = 1, \dots , 8)$ are supplemented by $\lambda^0=\lambda_0=\sqrt{2/3}$ times the $3\times 3$ unit matrix. The second term in the first line introduces additional vector and axial-vector interactions. Their general form, invariant under chiral ${\rm SU}(3)_{\rm L}\times{\rm SU}(3)_{\rm R}$ symmetry, is \cite{Klimt:1989pm,Vogl:1991qt}:
	\begin{equation*}
		\begin{aligned}
	\mathcal{L}_v&=-\dfrac{1}{2}\, g\sum_{a=1}^8(\bar q\,\gamma^\mu\lambda^a q)^2-\dfrac{1}{2}\,g_{v,0}\left(\bar q\,\gamma^\mu\lambda_0\, q\right)^2\\&\quad-\dfrac{1}{2}\, g\sum_{a=1}^8(\bar q\,\gamma^\mu\gamma_5\lambda^a q)^2-\dfrac{1}{2}\,g_{a,0}\left(\bar q\,\gamma^\mu\gamma_5\lambda_0 \,q\right)^2\,.
	\end{aligned}
	\end{equation*}
Using vector dominance and the small difference between the masses of $\rho$ and $\omega$ mesons, one can choose \cite{Klimt:1989pm,Klimt:1990ws} $g_{v,0}=g_{a,0}\equiv g$. In the following we work with a simplified ansatz keeping only the single term, 
	\begin{equation}
			\mathcal{L}_v\to-\dfrac{1}{2}\,G_v\left(\bar q\,\gamma^\mu q\right)^2\,,
		\end{equation}
with vector-coupling strength $G_v = {2\over 3}g$. If a color current-current interaction is chosen to start with, a Fierz transformation would relate the vector and scalar couplings as $G_v=\frac{1}{2}\,G$.

The term in the second line of Eq.\,\eqref{3flint} is the Kobayashi-Maskawa-'tHooft determinant \cite{PhysRevLett.37.8,Kobayashi:1973fv} that describes the (anomalous) breaking of the axial ${\rm U}(1)_{\rm A}$ symmetry and gives rise to the large mass of the $\eta'$ meson.
 
The PNJL model is non-renormalizable. It operates with a characteristic three-momentum cutoff scale $\Lambda$, such that the effective interaction between quarks is ``turned off" for  momenta $|\vec{p}\, | > \Lambda$. No additional divergences appear at finite temperature and density. We adopt the cutoff prescription given in Ref.~\cite{Bratovic:2012qs} and use the following parameters \cite{Klimt:1990ws}: $m_u=m_d=3.6\,{\rm MeV}$, $m_s=87.0\,{\rm MeV}$, $\varLambda=750\,{\rm MeV}$, $G=3.64/\varLambda^2$, $K=8.9/\varLambda^5$. With this parameter set the empirical meson spectrum and the measured pseudoscalar decay constants in vacuum are well reproduced. The value of the vector coupling strength, $G_v$, is varied in order to investigate the impact of the repulsive vector interaction on the equation of state. A study comparing various parameter sets 
within a similar framework is presented in \cite{Masuda:2012ed}.

In Eq.\,(\ref{spnjl}) the color gauge covariant derivative $D^\nu=\partial^\nu+\imu A^\nu=\partial^\nu+\imu\,\delta_0^\nu A^{0,a}\frac{\lambda_a}{2}$ involves the $\text{SU}(3)_c$ Gell-Mann matrices $\lambda_a,a\in\{1,\dots,8\}$. The gauge coupling is absorbed in the definition of $A^{0,a}$. The temporal gauge field $A^0$ is treated as a constant Euclidean background field in the form $A_4=\imu A^0=A_4^3\frac{\lambda_3}{2}+A_4^8\frac{\lambda_8}{2}$. The last term in Eq.~\eqref{spnjl} is the Polyakov-loop effective potential $\mathcal{U}$, multiplied by the volume $V$ and the inverse temperature $\beta=T^{-1}$, and constructed as follows:
	\begin{equation}\label{upolyakov}
		\begin{aligned}
		&\dfrac{\mathcal{U}(\Phi,\bar\Phi;T)}{T^4}=-\dfrac{1}{2}\,b_2(T)\,\Phi\bar\Phi\\
       &+b_4(T)\,\ln\left[1-6\Phi\bar\Phi+4(\Phi^3+\bar\Phi^3)-3(\Phi\bar\Phi)^2\right]\,
	\end{aligned}
	\end{equation}
where $\Phi$ and $\bar\Phi$ are represented as
	\begin{equation}
		\begin{aligned}
	\Phi&=\dfrac{1}{3}\left[\euler^{\imu\frac{A_4^3+A_4^8}{2T}}+\euler^{-\imu\frac{A_4^3-A_4^8}{2T}}+\euler^{\imu\frac{A_4^8}{\sqrt{3}T}}\right]\\
	\bar\Phi&=\Phi^*\,.
		\end{aligned}
	\end{equation}
The coefficients $b_2(T)$ and $b_4(T)$ are parametrized to reproduce pure-gauge lattice QCD results (cf.~Refs.~\cite{Rossner:2007ik,Roessner:2006xn,Hell:2008cc,Hell:2011ic}). The temperature $T_0$ appearing in $b_2(T)$ and $b_4(T)$ is set to the transition temperature for the confinement-deconfinement crossover in the presence of two light and one heavy quark, as discussed in Ref.~\cite{Schaefer:2007pw}.

Given this input, the grand-canonical potential $\Omega = - \ln\,{\cal Z}$ is calculated in mean-field approximation with the partition function ${\cal Z}$ constructed from the  action $\mathcal{S}_\text{PNJL}$ of Eq.\,\eqref{spnjl}. Details are relegated to the Appendix. The result is the thermodynamic potential $\Omega_{\text{MF}}$ given in Eq.\,(\ref{omegapnjl}). It involves the expectation values  of the scalar fields, $\bar \sigma_i = -G\langle\bar{q}_i q_i\rangle$ ($i\in\{u,d,s\}$) representing the chiral condensates for each quark species, and of the vector field, $\bar v = G_v\langle q^\dagger q\rangle$, related to the baryon number density  of the quarks.

Minimization of $\Omega_{\text{MF}}$ determines the fields $\bar\sigma_i$, $\bar v$, $A_4^3$, and $A_4^8$ from the set of equations
\begin{equation}\label{gaps}
	\dfrac{\partial\varOmega_\text{MF}}{\partial \bar\sigma_i}=\dfrac{\partial\varOmega_\text{MF}}{\partial \bar v}=\dfrac{\partial\varOmega_\text{MF}}{\partial A_4^3}=\dfrac{\partial\varOmega_\text{MF}}{\partial A_4^8}=0\,.
	\end{equation}
In particular, dynamical quark masses emerge from the gap equations (\ref{gapeqs}).
In mean-field approximation it follows that $\Phi=\bar\Phi$ and consequently $A_4^8=0$  as shown in Refs.~\cite{Roessner:2006xn,Rossner:2007ik}. In the limit $T\rightarrow 0$ one actually has $\Phi=\bar\Phi=0$.

With the aim of describing charge-neutral matter in chemical equilibrium inside neutron stars, the equations \eqref{gaps} have to be supplemented by the following conditions for the densities and chemical potentials of the quarks and leptons involved:
		\begin{align}
			\dfrac{2}{3}\,\varrho_u-\dfrac{1}{3}\,\varrho_d-\dfrac{1}{3}\,\varrho_s-\varrho_e-\varrho_\mu&=0\,,\label{chargeneutral}\\
			\mu_d=\mu_u+\mu_e\,,\qquad\mu_d=\mu_s\,,\qquad\mu_e&=\mu_\mu\,.\label{pnjlbeta}
	\end{align}
Eq.~\eqref{chargeneutral} expresses charge neutrality when both electrons and muons participate in etablishing chemical (beta) equilibrium. The particle densities are calculated from
	\begin{equation}\label{particledensity}
		\varrho_i=-\left(\dfrac{\partial\varOmega}{\partial\mu_i}\right)_{T,V,\{\mu_j\}_{j\neq i}}\,.
		\end{equation}
For the particle densities of the leptons $e,\mu$ we simply use those derived from the thermodynamic potential, $\varOmega_{\rm lepton}$, of a free gas of electrons and muons.
Beta equilibrium in terms of the processes 
\begin{align}
d&\leftrightarrow u+e^-+\bar\nu_e~,~~~~~ s\leftrightarrow u+e^-+\bar\nu_e~,\nonumber\\ d&\leftrightarrow u+\mu^-+\bar\nu_\mu~,~~~~s\leftrightarrow u+\mu^-+\bar\nu_\mu~,\nonumber 
\end{align}
is expressed by Eqs.~\eqref{pnjlbeta} (neglecting chemical potentials for neutrinos).

Consider now the EoS for beta-equilibrated quark matter. The gap equations \eqref{gaps} are solved simultaneously under the constraints of charge neutrality \eqref{chargeneutral} and beta equilibrium \eqref{pnjlbeta}. Only one of the chemical potentials remains as a free parameter. With the mean-field thermodynamic potential $\varOmega_{MF}$, the pressure of the 
system is 
	\begin{equation}
		P=-\varOmega_{\rm MF}-\varOmega_{\rm lepton}\, .
	\end{equation}
The energy density is calculated using the Gibbs-Duhem relation,
	\begin{equation}\label{gibbsduhem}
		\epsilon=T s-P+\sum_i\mu_i\varrho_i\,,
	\end{equation}
where the particle densities, $\varrho_i$, are given in Eq.~\eqref{particledensity}, and the entropy density $s$ is determined as
\begin{equation}
	s=-\left({\partial\Omega\over\partial T}\right)_{V,\{\mu_j\}	}\, .
\end{equation}

Resulting equations of state at $T=0$ are shown in Fig.~\ref{EoS} for different values of the vector coupling strength $G_v$. It actually turns out that low temperatures $T\lesssim 10\,{\rm MeV}$ do not affect the EoS for $\varrho=\frac{1}{3}(\varrho_u+\varrho_d+\varrho_s)\gtrsim \varrho_0$. In what follows we use $T=0$ throughout. Fig.~\ref{EoS} displays a qualitative change in the properties of the EoS, depending sensitively on the vector coupling strength. For $G_v=0$ the low-temperature EoS features a first-order chiral phase transition leading to an EoS that is far too soft and fails to satisfy the neutron star constraints. This first-order transition disappears and turns into a continuous crossover once the repulsive vector interaction strength exceeds a critical value, $G_v^{\text{crit}}\simeq 0.9\,G$. The constraints from neutron star observables would require a further strengthening of the vector repulsion between quarks, up to $G_v\simeq 1.5\,G$ as demonstrated in Fig.~\ref{EoS}.

\begin{figure}[htbp]
\begin{center}
 \includegraphics[width=.48\textwidth]{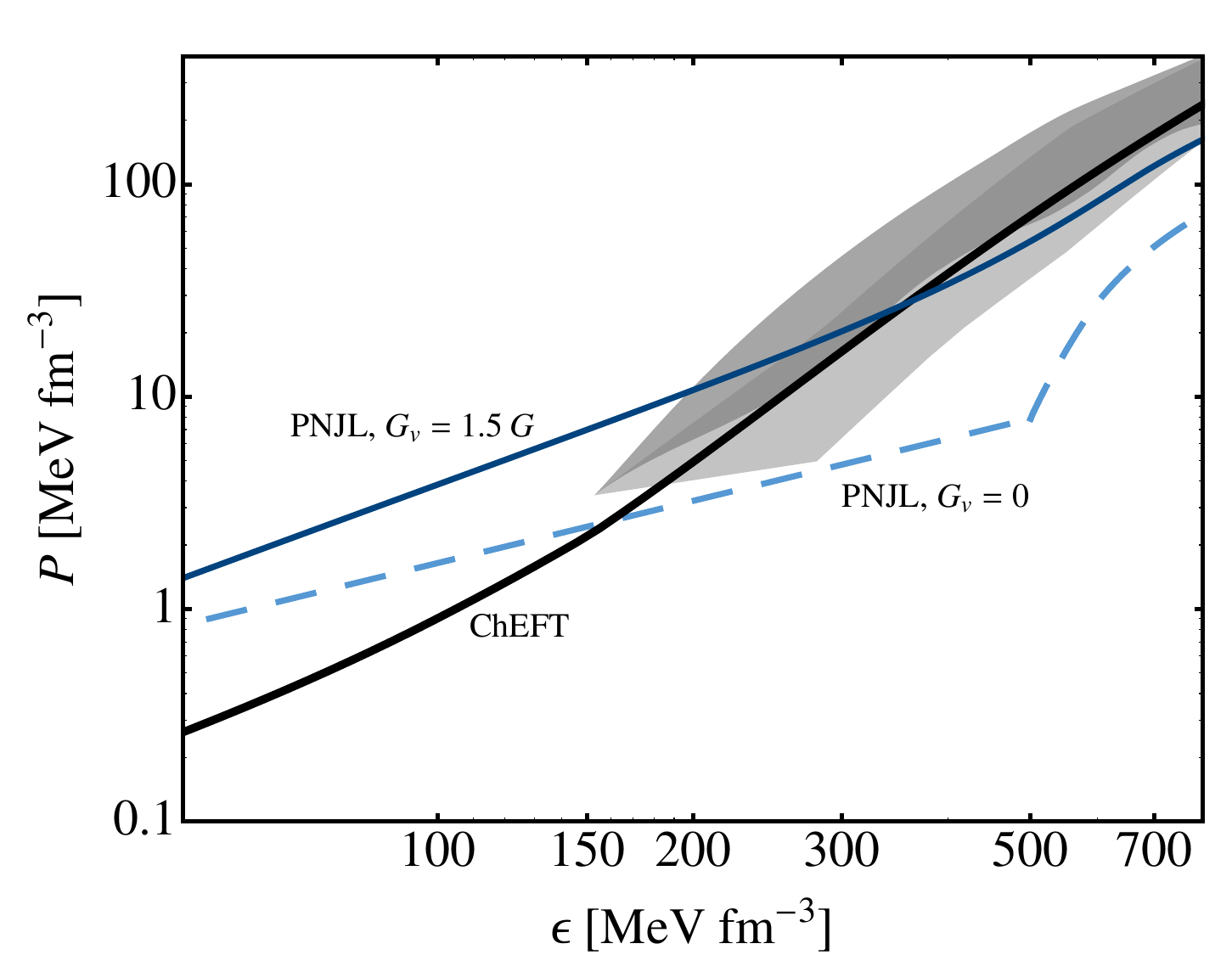}
\end{center}
\vspace{-.5cm}
\caption{Equations of state at $T = 0$ derived from the 3-flavor PNJL model with inclusion of charge neutrality and beta equilibrium conditions. The blue dashed and solid lines show results for different vector coupling strengths $G_v$ as indicated in the figure. The black solid line displays the EoS derived from in-medium chiral effective field theory as described in the previous section \ref{chefteos} and discussed further in the next section. The grey bands show the constraints from neutron star observables (see Fig.\,\ref{constraintseos}). }
\label{EoS}
\end{figure}

It is instructive to study the particle ratios,
\begin{equation}
\frac{\varrho_i}{\varrho_u+\varrho_d+\varrho_s}~~~~~~ (i =u,d,s,e)~~,\nonumber		
	\end{equation}
as they emerge from this (P)NJL model, as a function of the baryon density
	\begin{equation}
		\varrho=\dfrac{1}{3}\left(\varrho_u+\varrho_d+\varrho_s\right) ~.\nonumber
		\end{equation}
These particle ratios turn out to be universal: they do not depend on the strength of the vector interaction. This is because the vector field, $\bar v$, appears only in the combination $\mu_i-\bar v$ with the chemical potentials $\mu_i$. 
\begin{figure}[htbp]
\begin{center}
 \includegraphics[width=.45\textwidth]{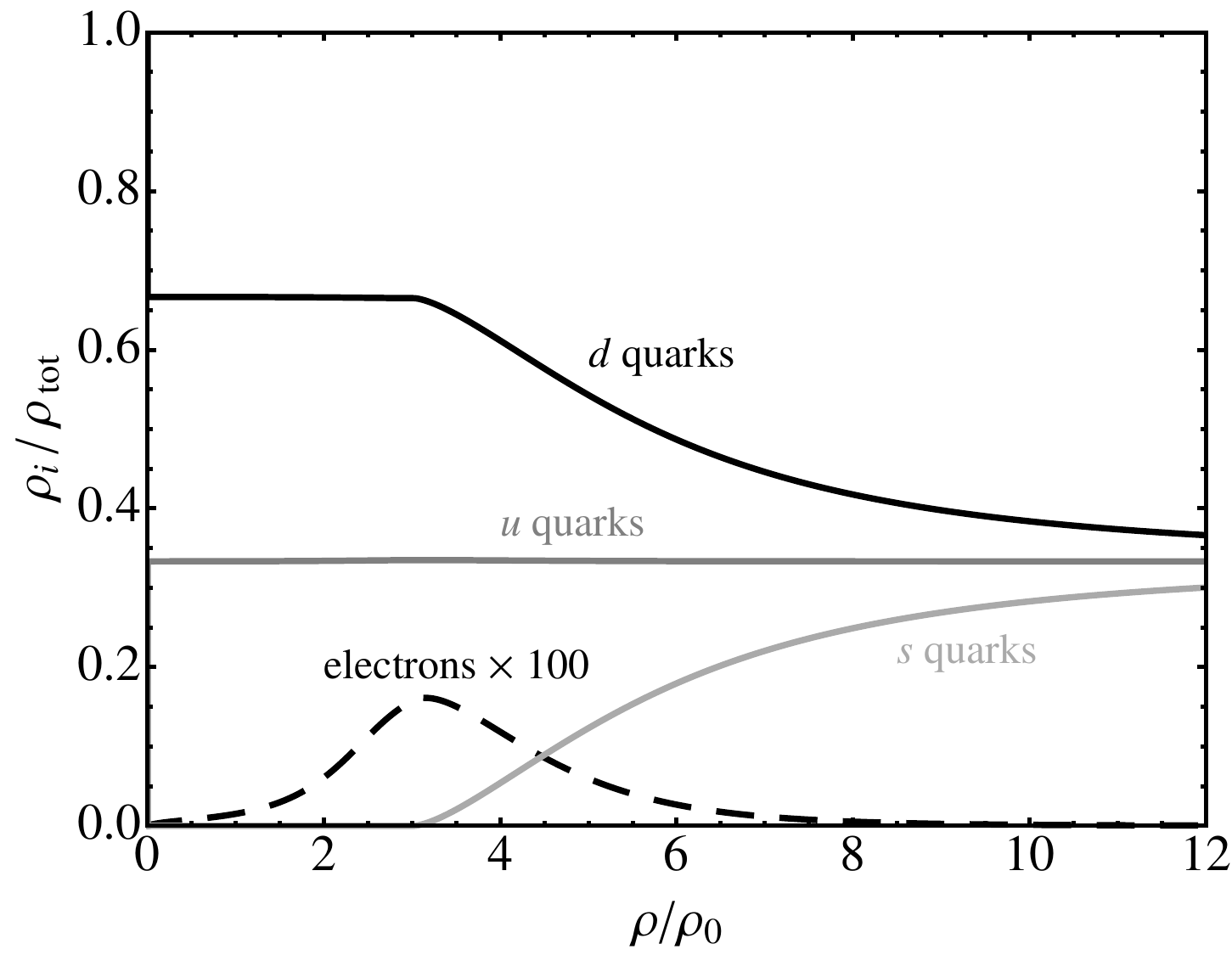}
\end{center}
\vspace{-.5cm}
\caption{Particle ratios in the 3-flavor (P)NJL model subject to beta equilibrium and charge neutrality. The ratios $\varrho_i /\varrho_{\rm tot}$ with $\varrho_{\rm tot} = \varrho_u + \varrho_d + \varrho_s$, for the species indicated in the figure, are given as a function of the baryon density (normalized to nuclear saturation density $\varrho_0=0.16\,{\rm fm}^{-3}$).
}
\label{pnjlparticleratiosfig}
\end{figure}
The result is shown in Fig.~\ref{pnjlparticleratiosfig}. The muon fraction is always zero because the muon chemical potential never exceeds the muon mass. At low baryon densities, $\varrho\lesssim 3\,\varrho_0$ with $\varrho_0=0.16\,{\rm fm}^{-3}$, the relative proportion of $d$ and $u$ quarks is reminiscent of neutron matter. At densities $\varrho\gtrsim4\,\varrho_0$ strange quarks start to become important. As will be pointed out in Sec.~\ref{neutronstarsection} such densities can only be reached at the very center of the inner core in neutron stars. At these densities the strange-quark chemical potential $\mu_s$ exceeds the constituent quark mass $M_s$. Given their negative charge, strange quarks can now replace the electrons and at the same time reduce the fraction of $d$ quarks.  At $\varrho\gtrsim10\varrho_0$ the densities of all three quark species approach each other, indicating that the quarks become flavor degenerate at the highest densities. 

\end{subsection}
\end{section}

\begin{section} {Nucleonic versus hybrid equations~of~state}
\label{neutronstarsection}

Given the equations of state for different realizations of dense baryonic matter as derived in Sec.~\ref{eossection}, we now proceed with a discussion of several scenarios, ranging from a purely nucleonic composition to hybrid hadron-quark matter, always subject to the constraints provided by neutron star observables and presented in Sec.~\ref{massradiusconstraintssection}. 

\begin{subsection}{Conventional nuclear matter}\label{nsnuclearmattersection}

Consider first the  EoS based entirely on nuclear chiral effective field theory as described in Sec.~\ref{chefteos}. We recall that this EoS is generated using in-medium chiral perturbation theory to three-loop order in the energy density. It includes explicitly one- and two-pion exchange dynamics and three-body forces in the presence of the nuclear medium, together with re-summed contact terms. The energy density is written as:
\begin{equation}
		\epsilon(\varrho,x_p)=\varrho\left[M_N+\bar E(\varrho,x_p)\right]\,,
	\end{equation}
with the energy per nucleon, $\bar{E} = E/A$, given as a function of the density $\varrho = \varrho_n + \varrho_p$ and the proton fraction, $x_p = \varrho_p/\varrho$. The expansion of $\bar{E}$ provided by in-medium chiral effective field theory is actually in powers of the Fermi momentum, i.e. in fractional powers of the density $\varrho$. The nucleon mass is taken as the average of neutron and 
proton masses, $M_N=\frac{1}{2}\left(M_n+M_p\right)$. As mentioned previously it is useful to write the energy per nucleon as an expression to second order in the asymmetry parameter, $\delta=(\varrho_n-\varrho_p)/\varrho$, given the small proton fraction $x_p$ encountered in the neutron star interior. With the calculated energies per nucleon for symmetric nuclear matter, $\bar{E}_{SM}$, and pure neutron matter, $\bar{E}_{NM}$, and the symmetry energy,
$S(\varrho) = \bar{E}_{NM}(\varrho) - \bar{E}_{SM}(\varrho)$:
\begin{equation}\label{easymexpansion}
		\begin{aligned}
		\bar E&=\bar E_{SM}(\varrho)+S(\varrho)(1-2 x_p)^2
			\\&=\left(1-2 x_p\right)^2\bar E_{NM}(\varrho) + 4 x_p(1-x_p)\,\bar E_{SM}(\varrho)\,.
		\end{aligned}
	\end{equation}
The ChEFT calculation of the symmetry energy at nuclear saturation density, $\varrho_0=0.16$ fm$^{-3}$, gives 
\begin{equation}
		S_{\rm ChEFT}(\varrho_0)=33.5\,{\rm MeV}\,,
	\end{equation}
compatible with empirically deduced values that range between 26 and 44 MeV \cite{Chen:2007ih}. It is common to expand the symmetry energy around nuclear saturation density,
	\begin{equation}
		S(\varrho)=S(\varrho_0)+\dfrac{L}{3}\left(\dfrac{\varrho-\varrho_0}{\varrho_0}\right)+\dots
	\end{equation}
The $L$ value,
		\begin{equation}
		L=\left. 3\varrho_0\dfrac{\partial S}{\partial \varrho}\right|_{\varrho=\varrho_0}\,,
		\end{equation}
is poorly known and supposed to be in the range 50 MeV $\lesssim L \lesssim$ 140 MeV (see \cite{Chen:2007ih,Li:2008gp} and references therein). Our calculation gives
	\begin{equation}
		L_{\rm ChEFT}=48\,{\rm MeV}\,,
	\end{equation}
at the lower side of the empirical bandwidth. The significance of the $L$ value is that it scales linearly with the neutron-skin thickness (i.e., the difference between the root-mean-square radii of neutron and proton distributions) of heavy nuclei \cite{Furnstahl:2001un}. Implications of the symmetry energy for neutron stars are discussed in Ref.~\cite{Danielewicz:2002pu}.

Beta equilibrium involving electrons and muons, $n\leftrightarrow p+e^-+\bar\nu_e$ and $n\leftrightarrow p+\mu^-+\bar\nu_\mu$, together with charge neutrality imply:
	\begin{align}
			\varrho_p &= \varrho_e +\varrho_\mu\,,\label{cheftchargeneutral}\\
			\mu_n= \mu_p&+\mu_e\,,\quad\mu_e=\mu_\mu\,.\label{cheftbeta}
		\end{align}
where the neutron and proton chemical potentials are given by 
\begin{equation}
\mu_{n,p} = \left({\partial\epsilon\over\partial\varrho_{n,p}}\right)_V~.
\end{equation}
The lepton charge densities, $\varrho_e,\varrho_\mu$, and the corresponding chemical potentials, $\mu_e,\mu_\mu$, are again assumed to be those of a free Fermi gas of electrons and muons. 

Incorporating the conditions \eqref{cheftchargeneutral} and \eqref{cheftbeta} the equation of state $P(\epsilon)$, applicable for neutron star matter in beta equilibrium at zero temperature, is derived using
\begin{equation}
		P = -\epsilon+\sum_i\mu_i\varrho_i\,~~~~ (i = n,p;e,\mu)~.
	\end{equation} 
At very low densities this EoS based on ChEFT is matched again to the ``SLy'' EoS as in Fig.\,\ref{constraintseos}. The complete result is shown by the solid black curve in Fig.\,\ref{EoS}. Evidently the ChEFT equation of state satisfies the astrophysical constraints over the whole range of relevant energy densities. The exact microscopic treatment of the Pauli principle acting on the in-medium pion-exchange processes and the repulsive three-nucleon correlations provide the required stiffness of the EoS in the dense medium to support two-solar-mass neutron stars.

\begin{figure}[htbp]
\begin{center}
 \includegraphics[width=.4\textwidth]{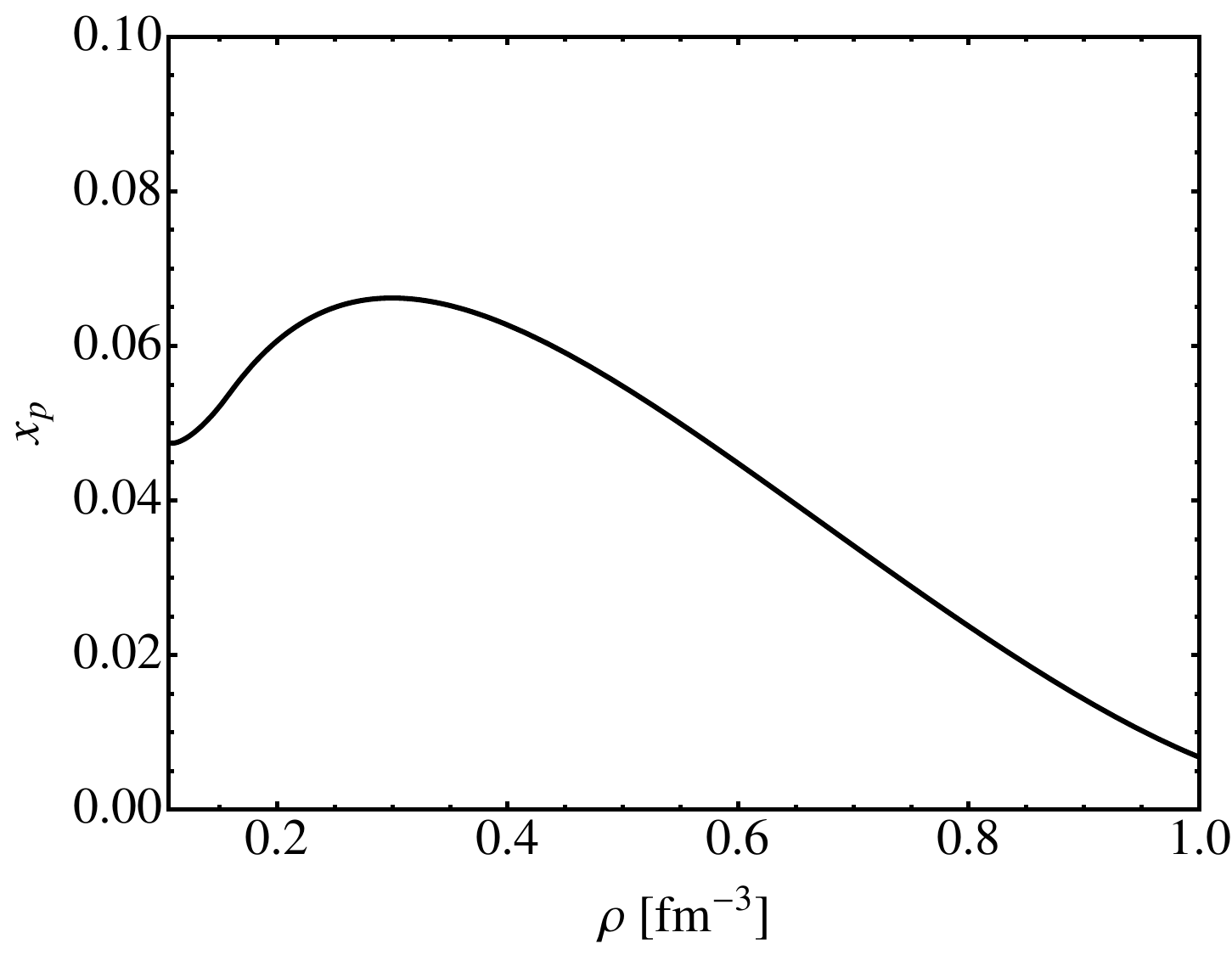}
\end{center}
\vspace{-.5cm}
\caption{Proton fraction, $x_p=\frac{\varrho_p}{\varrho}$, shown for the ChEFT EoS including beta equilibrium. 
}
\label{xpcheftfig}
\end{figure}

The proton fraction $x_p$ in neutron star matter follows from the ChEFT equation of state is shown in Fig. \ref{xpcheftfig}. The smallness of the proton admixture (which stays systematically below a maximum of less than 7\,\% reached at about twice the density of normal nuclear matter) justifies the ansatz quadratic in $x_p$ as written in Eq.\,(\ref{easymexpansion}).

\begin{figure}[htbp]
\begin{center}
 \includegraphics[width=.45\textwidth]{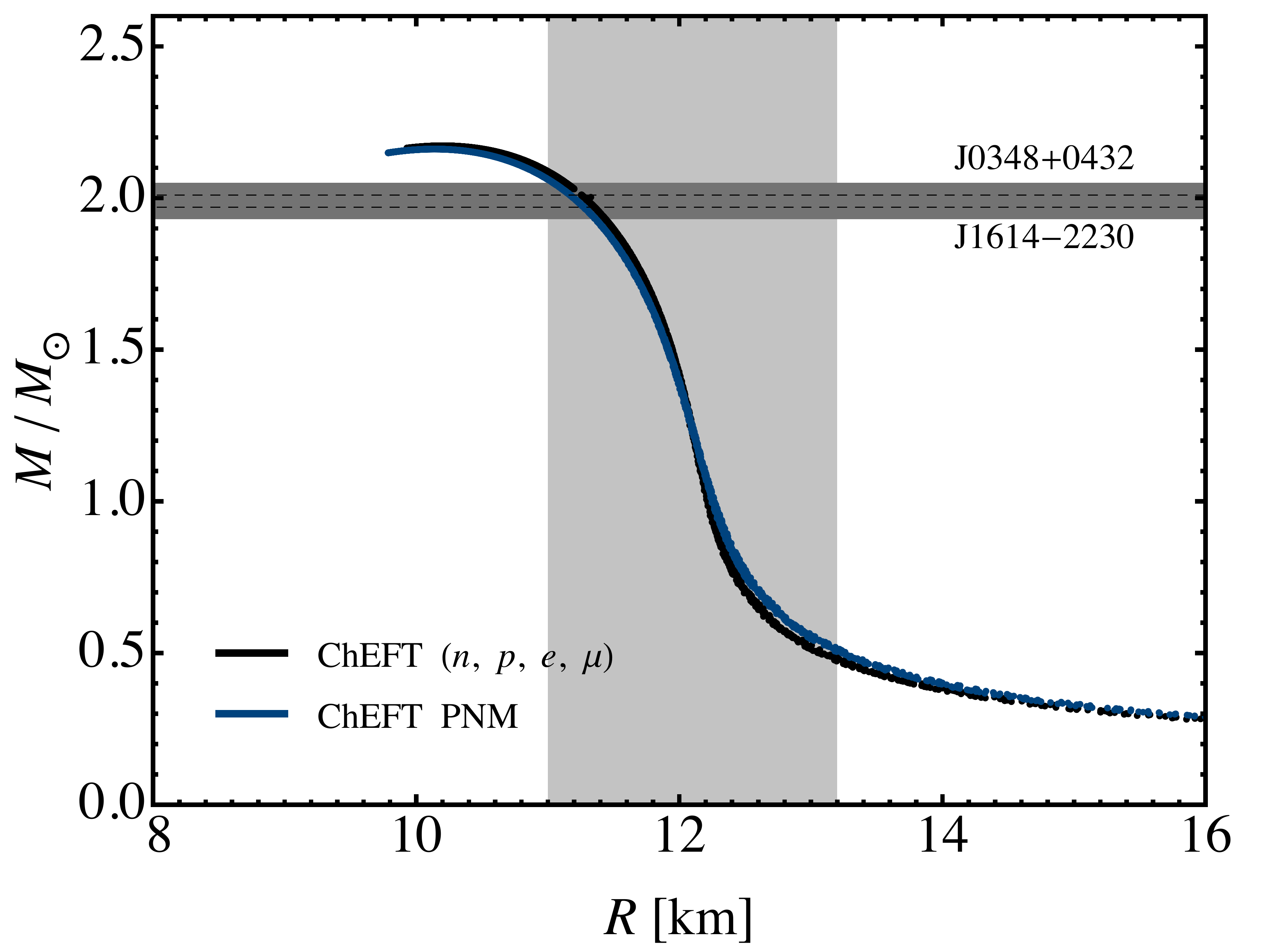}
\end{center}
\vspace{-.5cm}
\caption{Mass-radius relation computed with the ChEFT equation of state for neutron stars including beta equilibrium. Stable neutron stars can exist up to the maximum of this curve. The hardly distinguishable EoS for pure neutron matter (PNM) is also shown for reference. The horizontal band indicates the masses of the pulsars J1614--2230 and J0348+0432. The lighter grey band corresponds to the radius range deduced in Ref.\,\cite{Steiner:2010fz}.
}
\label{cheftmassradiusfig}
\end{figure}

Given the pressure as a function of energy density the TOV equations \eqref{tov} and \eqref{massdgl} are solved. The resulting ChEFT mass-radius relation for neutron stars is shown in Fig.\,\ref{cheftmassradiusfig}. It turns out that there is only a marginal difference between the results for pure neutron matter and matter in beta equilibrium with its small proton admixture. In either case the equation of state is sufficiently stiff to pass beyond the two-solar-mass threshold. Our results are compatible with the accepted range of neutron star radii according to Ref.\,\cite{Steiner:2010fz} and also (within limits) of Ref.\,\cite{Truemper2013}. 

\begin{figure}[htbp]
\begin{center}
 \includegraphics[width=.4\textwidth]{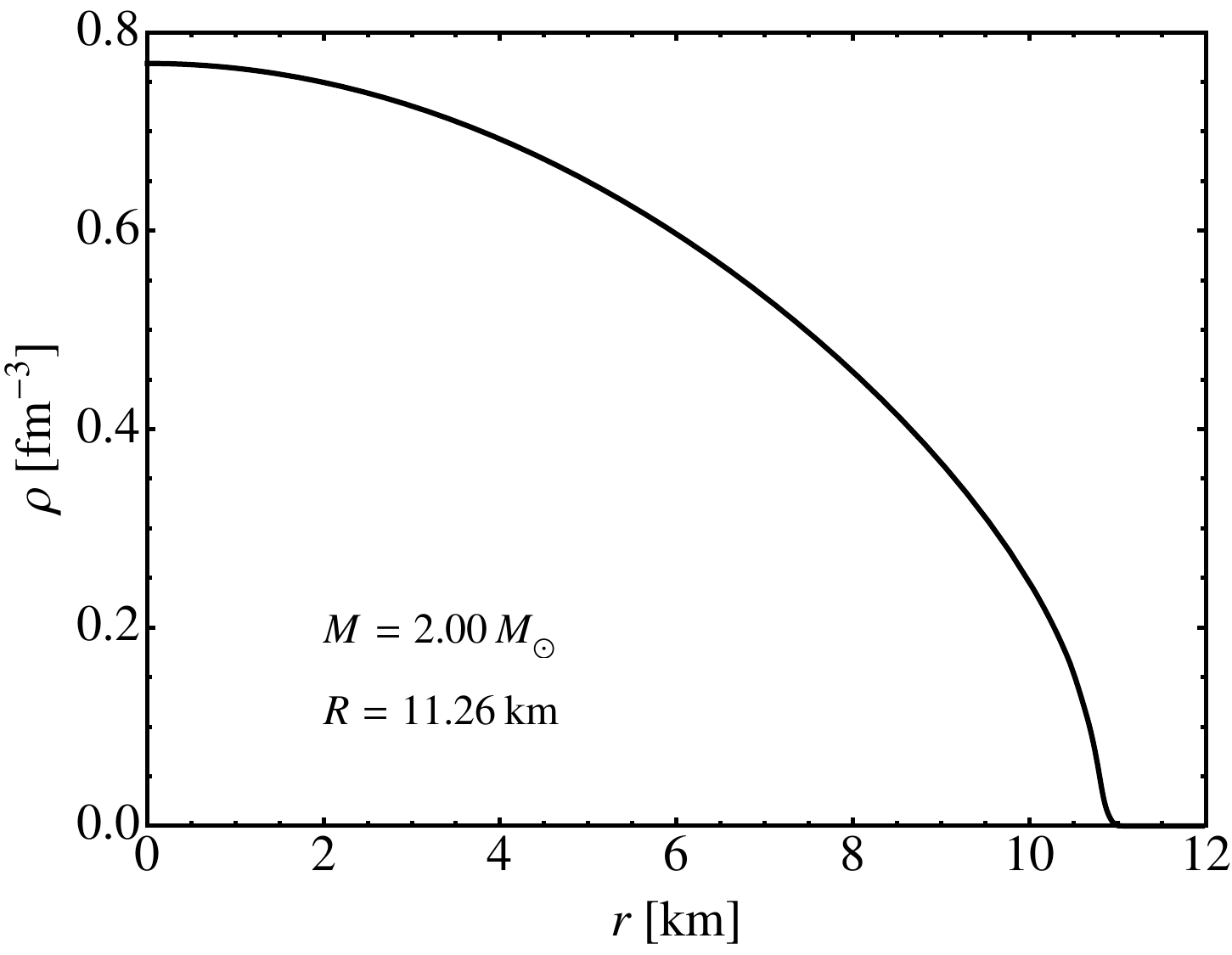}
\end{center}
\vspace{-.5cm}
\caption{Density profile in the interior of a neutron star with mass $M=2\,M_\odot$ and resulting radius of about $R=11\,{\rm km}$. Note that the central density does not exceed $\varrho_c\sim4.8\,\varrho_0$.
}
\label{chefttsmnsfig}
\end{figure}

Next, consider the calculated density profile of a neutron star with a mass $M=2\,M_\odot$, displayed in  Fig.~\ref{chefttsmnsfig}. As a general feature of a stiff EoS, the baryon density $\varrho_c$ reached in the center of the star is by far lower than the values characteristically associated with many previous neutron star models which worked with softer equations of state.  In the present example, the central density does not exceed about $\varrho_c\simeq 4.8\,\varrho_0$. Such bounds on the central density are also characteristic of advanced calculations using quantum Monte Carlo techniques \cite{GCR:2012}.

Concerns might still be raised about how far ChEFT calculations can be extrapolated into the high-density regime. A necessary condition for the applicability of in-medium ChEFT is that the medium persists in the hadronic phase of QCD with spontaneously broken chiral symmetry. Investigations of the in-medium chiral condensate at zero temperature \cite{FKW:2012, HKW:2013,Drews:2013,Drews:2014} do indeed show a stabilization of the density-dependent condensate $\langle\bar{q}q\rangle(\varrho, T = 0)$, shifting the transition to chiral symmetry restoration far beyond three times $\varrho_0$. 

The in-medium ChEFT approach relies on the assumption that the proper baryonic degrees of freedom are nucleons (rather than liberated quarks) even in compressed baryonic matter. In this context the following qualitative picture may be useful for orientation.
Models based on the chiral symmetry of QCD describe the nucleon \cite{TW:2001} as a compact valence quark core with a radius of about 1/2 fm, surrounded by a pionic cloud. The meson cloud determines most of the empirical proton rms charge radius of 0.87 fm. For the neutron the picture of core and cloud is analogous except that the electric charges of quark core and meson cloud now add up to form the overall neutral object. Even at $\varrho \sim 5\,\varrho_0$ the typical average distance between two neutrons is about 1 fm, hence the baryonic cores still do not overlap appreciably at such densities. The pionic field surrounding the baryonic sources is of course expected to be highly inhomogeneous and polarized in compressed matter, but this effect is properly dealt with in chiral EFT. It is therefore perhaps not so surprising that an EoS based entirely on nucleons (plus $\Delta$ isobars) and pionic degrees of freedom works well for neutron stars, once the repulsive mechanisms for generating stiffness and high pressure are properly incorporated. A similiar reasoning is found e.g. in Ref.\,\cite{Pandharipande:1998bs}.

\end{subsection}

\begin{subsection}{Hybrid stars}

This subsection deals with the possibilty that the inner core of the neutron star is composed of quark matter. It is obviously not realistic to think of a quark matter EoS for the entire core region. But a combination of a suitable quark matter equation of state for the inner core with the ChEFT EoS from the previous section, describing the outer core, is still an option. In the following we discuss two scenarios: first an ansatz featuring quark-hadron continuity, and secondly a first-order phase transition involving a coexistence region of hadronic and quark matter.

\begin{subsubsection}{Quark-hadron continuity}

The quark-hadron continuity picture has been discussed previously in Refs.\,\cite{Schafer:1998ef, 
Masuda:2012ed, fukushima-2004-591,Baym:2008me,Maeda:2009ev}. It is based on the assumption that the outer and inner core regions of the hybrid neutron star are characterized by a smooth, continuous transition between the nucleonic and quark matter regions. 

Hybrid scenarios were also studied in Ref.\,\cite{Alford:2005} where it was pointed out that the appearances of ordinary neutron stars and hybrid stars can be quite similar. Hybrid stars with  hyperons and including effects of quark color super-conductivity were explored in Ref.\,\cite{Bonanno:2012} (not respecting, however, the nuclear physics constraints emphasized in the present work).

Here we follow an ansatz introduced in Ref.\,\cite{Masuda:2012ed}  and combine the ChEFT EoS representative of hadronic (nucleonic plus pionic) matter, $P_{\rm H}(\epsilon)$, with the quark matter EoS derived from the PNJL model, $P_{\rm Q}(\epsilon)$: 
	\begin{equation}\label{pnjlcontinuity}
		P(\epsilon)= P_{\rm H}(\epsilon) f_{\rm H}(\epsilon)+P_{\rm Q}(\epsilon) f_{\rm Q}(\epsilon)\,,
	\end{equation}
with interpolating functions:
	\begin{equation}\label{pnjlcontinuityf}
		\begin{aligned}
		 f_{\rm H}(\epsilon)&=\dfrac{1}{2}\left[1-\tanh\left(\dfrac{\epsilon-\bar\epsilon}{\varGamma}\right)\right]\,,\\
		 f_{\rm Q}(\epsilon)&=\dfrac{1}{2}\left[1+\tanh\left(\dfrac{\epsilon-\bar\epsilon}{\varGamma}\right)\right]\,.
	\end{aligned}
	\end{equation}
The parameters $\bar\epsilon$ and $\varGamma$ determine the location and the width of the transition region between the nucleonic and quark matter sectors. The pressure functions $P_{\rm H}(\epsilon)$ and $P_{\rm Q}(\epsilon)$ are matched continuously. The density, $\varrho=\varrho(\epsilon)$, can be determined from the EoS \eqref{pnjlcontinuity} by integrating 
\begin{equation}
{\diff\varrho\over\varrho}  = {\diff\epsilon\over P(\epsilon) + \epsilon} ~.
\end{equation}
	
In Fig.\,\ref{pnjlcontinuityeosfig} we show the EoS derived from Eq.\,\eqref{pnjlcontinuity} for different values of the NJL vector coupling strength $G_v$. We have chosen $\varGamma=300\,{\rm MeV}/{\rm fm}^3$ and $\bar\epsilon=800\,{\rm MeV}/{\rm fm}^3$, representing a transition region $3.0\,\varrho_0\lesssim\varrho\lesssim5.5\,\varrho_0$.  As in Sec.\,\ref{nsnuclearmattersection} the ``SLy" EoS has been matched smoothly to the ChEFT EoS at $\epsilon=100\,{\rm MeV}/{\rm fm}^3$. It is evident from the figure that a hadron-quark hybrid scenario meets the constraints from neutron star observables only if the repulsive vector coupling between quarks is sufficiently large, $G_v>G$.  This is confirmed by the mass-radius plot shown in Fig.\,\ref{pnjlcontinuitymrfig}. At this point our results are qualitatively similar to those of Ref.\,\cite{Masuda:2012ed} despite their use of a different hadronic EoS and of a different method.
	
\begin{figure}[htbp]
\begin{center}
 \includegraphics[width=.45\textwidth]{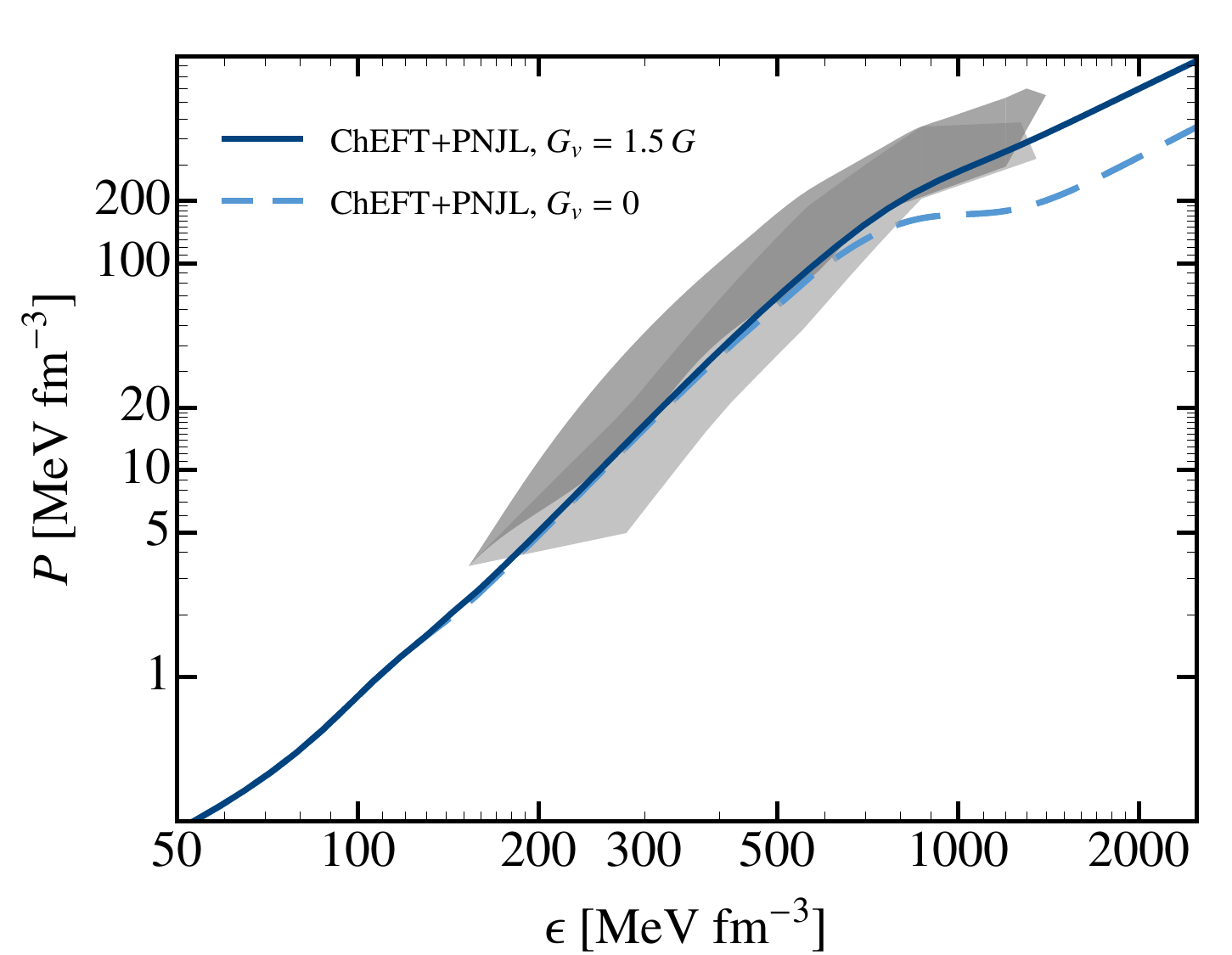}
\end{center}
\vspace{-.5cm}
\caption{Equations of state representing the quark-hadron continuity scenario using different quark vector couplings. Quark matter (PNJL) and nuclear matter (ChEFT) equations of state are matched continuously at $\epsilon=\bar\epsilon=800\,{\rm MeV}/{\rm fm}^3$. Solid curve: $G_v=1.5\,G$;  dashed curve: $G_v=0$.  The grey areas are those of Fig.\ref{constraintseos} representing constraints
from neutron star observables.
}
\label{pnjlcontinuityeosfig}
\end{figure}

\begin{figure}[htbp]
\begin{center}
 \includegraphics[width=.45\textwidth]{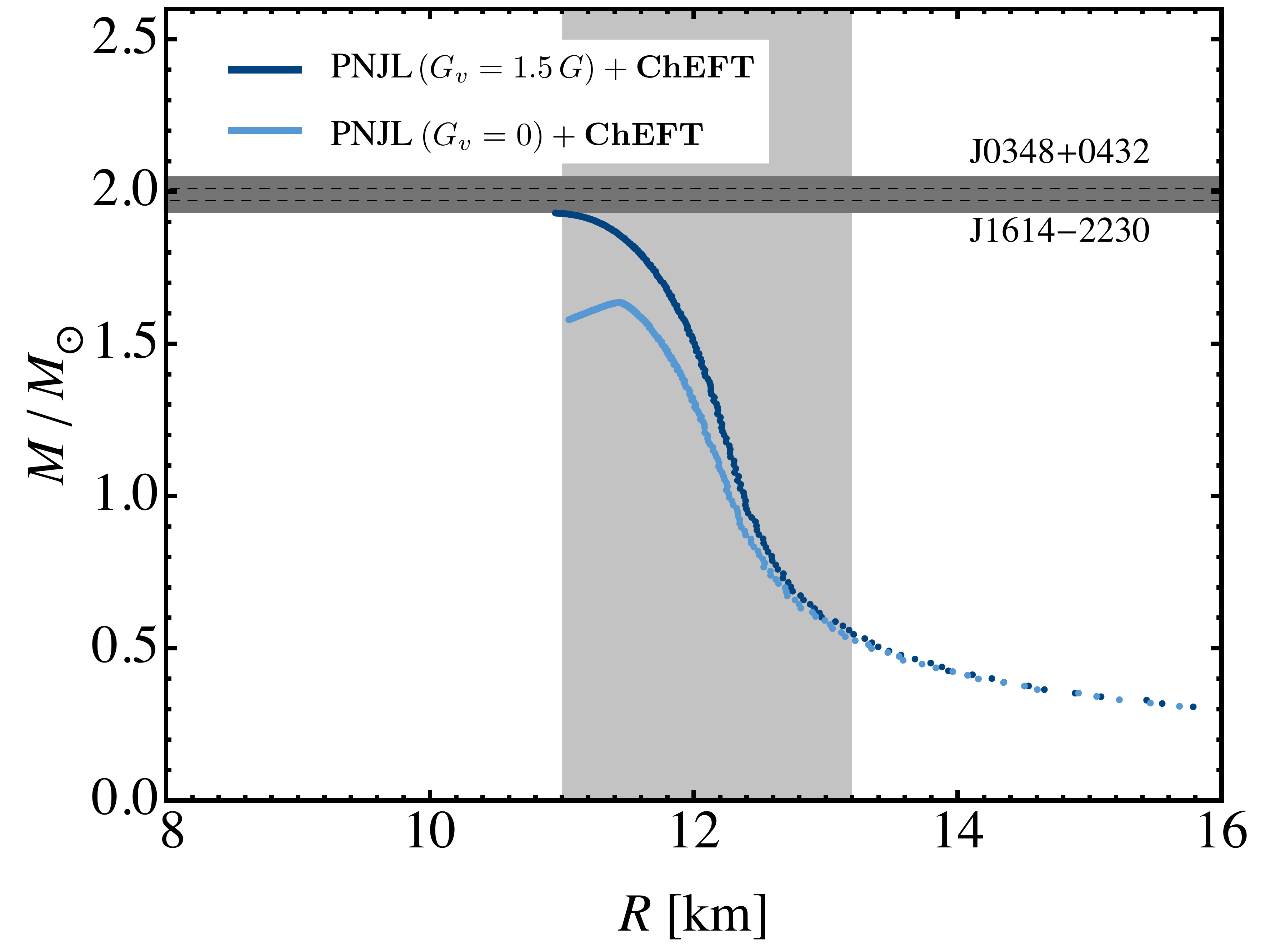}
\end{center}
\vspace{-.5cm}
\caption{Solutions of the TOV equations \eqref{tov} and \eqref{massdgl} (mass-radius relation) for neutron stars using the EoS given in \eqref{pnjlcontinuity}. The lines correspond to different vector coupling strengths, as indicated in the figure. The shaded areas are as in Fig.\,\ref{cheftmassradiusfig}.
}
\label{pnjlcontinuitymrfig}
\end{figure}

\end{subsubsection}

\begin{subsubsection}{Hadron-quark first-order phase transition}

In the previous section the transition region from the hadronic to the quark phase was chosen by means of the parameters $(\varGamma,\bar\epsilon)$ of the interpolating functions \eqref{pnjlcontinuityf}. In this section a different approach is taken assuming a first-order phase transition from hadronic to quark matter with an extended coexistence region of the two phases. The system is characterized by two conserved quantities: electric charge and baryon number. For such systems with more than one conserved charge, the Maxwell construction is generalized and replaced by the Gibbs condition \cite{Glendenning:1992vb}. In the present case, this condition describing mechanical and chemical equilibrium is:
	\begin{equation}\label{gibbs}
		P_{\rm H}(\mu_n,\mu_e)=P_{\rm Q}(\mu_n,\mu_e)~,
	\end{equation}
expressed as the pressure balance between hadronic and quark components in terms of the neutron and electron chemical potentials. For the nucleonic phase, the proton chemical potential is $\mu_p=\mu_n-\mu_e$. The muon chemical potential is $\mu_\mu=\mu_e$. For the quark matter phase, the quark chemical potentials are expressed in terms of $\mu_n$ and $\mu_e$ according to
	\begin{equation}
			\begin{aligned}
		\mu_u&=\dfrac{1}{3}(2\mu_p-\mu_n)=\dfrac{1}{3}(\mu_n-2\mu_e)\,,\\
		\mu_d&=\mu_s=\dfrac{1}{3}(2\mu_n -\mu_p)=\dfrac{1}{3}(\mu_n+\mu_e)\,.
	\end{aligned}
	\end{equation}
The choice of the chemical potentials $\mu_n,\mu_e$ is arbitrary. Note that $P_{\rm Q}$ also depends on the mean fields $\bar \sigma_i,\bar v$ which are in turn dependent on $\mu_n$ and $\mu_e$. The total baryon density in the coexistence region is \cite{Glendenning:1992vb}:
	\begin{equation}
		\varrho=\chi\,\varrho_{\rm Q}+(1-\chi)\,\varrho_{\rm H}\,,
		\end{equation}
where $\chi$ (with $0\le \chi \le 1$) denotes the proportion of quark matter in the hadron-quark
mixed system. The combinations $\chi\,\varrho_{\rm Q}$ and $(1-\chi)\,\varrho_{\rm H}$ are the densities of deconfined quarks and confined baryons, respectively, in the coexistence region. Global charge neutrality implies:
	\begin{equation}\label{gibbschargeneutral}
		\chi\sum_{i=u,d,s} q_i\,\varrho_i+(1-\chi)\,\varrho_p-\varrho_e-\varrho_\mu= 0\,,
	\end{equation}
where the $q_i$ denote the quark charges. 
	
The Gibbs condition \eqref{gibbs} together with Eq.\,\eqref{gibbschargeneutral} allows to eliminate two of the three quantities $\mu_n,\mu_e,\chi$. The pressure is a function of the remaining (free) parameter. The resulting equation of state is shown in Fig.\,\ref{gibbseosfig}. The corresponding particle densities (for the case $G_v=0$) are displayed in Fig.\,\ref{gibbsdensityfig}. The coexistence region in the case without vector interaction, $G_v=0$, extends over the baryon density interval $4\,\varrho_0\lesssim\varrho\lesssim 9\,\varrho_0$. For $G_v=0.5\,G$ the coexistence region is shifted to $6\,\varrho_0\lesssim\varrho\lesssim10\,\varrho_0$. Hence the phase transition takes place over a broad density range and moves toward higher densities as the vector repulsion is increased. An interesting feature observed in Fig.\,\ref{gibbseosfig} is the increase of the proton fraction to about 10\% in the coexistence region. This is primarily to compensate the increasing supply of negative charges from the emergent $d$ and $s$ quarks.

\begin{figure}[htbp]
\begin{center}
 \includegraphics[width=.45\textwidth]{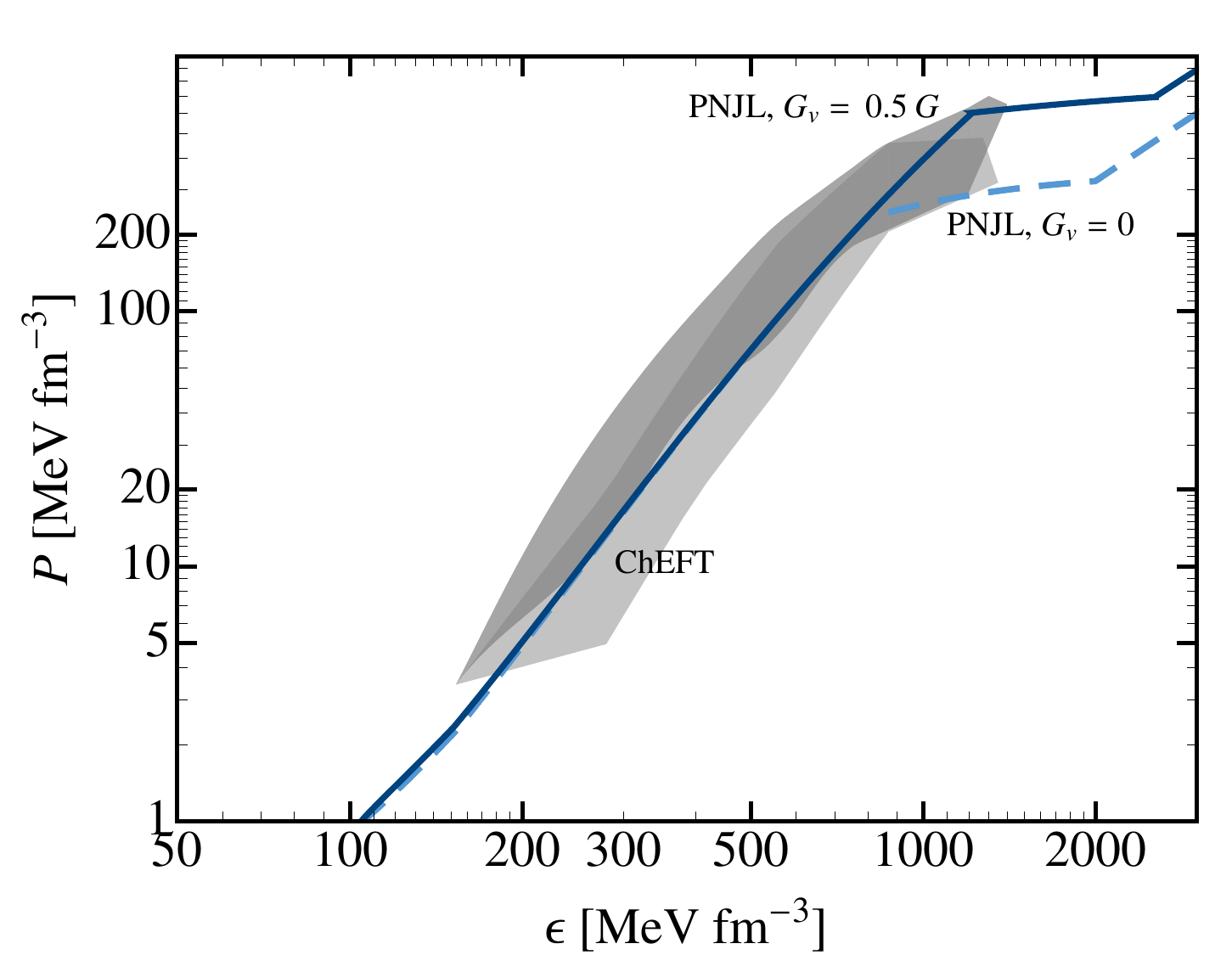}
\end{center}
\vspace{-.5cm}
\caption{Equations of state including a first-order phase transition between hadronic and quark matter. The transition region itself is characterized by the flat parts of the curves. The (upper) solid curve includes a vector repulsion of $G_v=0.5\,G$ between quarks, while the (lower) dashed curve is found using $G_v=0$. The grey areas are as in Fig.\,\ref{constraintseos}.
}
\label{gibbseosfig}
\end{figure}

\begin{figure}[htbp]
\begin{center}
 \includegraphics[width=.45\textwidth]{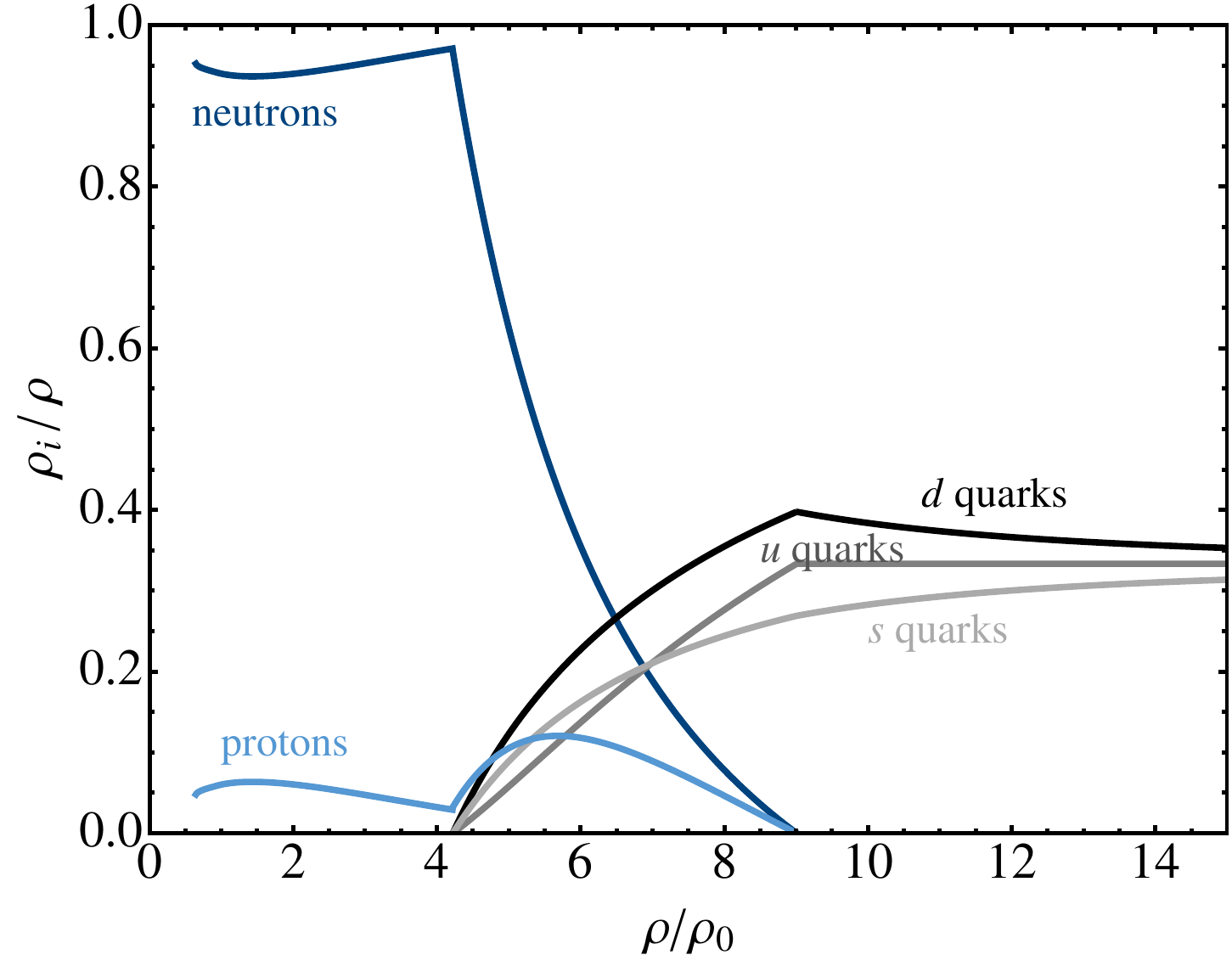}
\end{center}
\vspace{-.5cm}
\caption{Particle ratios as a function of the (normalized) baryon density for the particles as indicated in the figure. The first-order coexistence region is marked by the rapid decrease of neutrons and the steep rise of quarks. The case without vector interaction ($G_v=0$) is shown.
}
\label{gibbsdensityfig}
\end{figure}

The first-order quark-hadron transition softens the EoS. The impact of the phase transition is visible in the mass-radius plot of Fig.\,\ref{gibbsmrfig}. The $G_v=0$ case is interesting with its rapid turn of the mass-radius trajectory once the coexistence region is entered. While the two-solar-mass threshold is barely touched, the opening of the hadron-quark hybrid regime bends the $M(R)$ curve downward causing instability of the neutron star. Stability is recovered when the repulsive vector interaction between quarks is
introduced (with $G_v=0.5\,G$ in our example). However, in this case the first-order phase transition moves to densities $\varrho\gtrsim 6\,\varrho_0$, exceeding the maximal central density that can be realized in the inner core of the star. 

\begin{figure}[htbp]
\begin{center}
 \includegraphics[width=.45\textwidth]{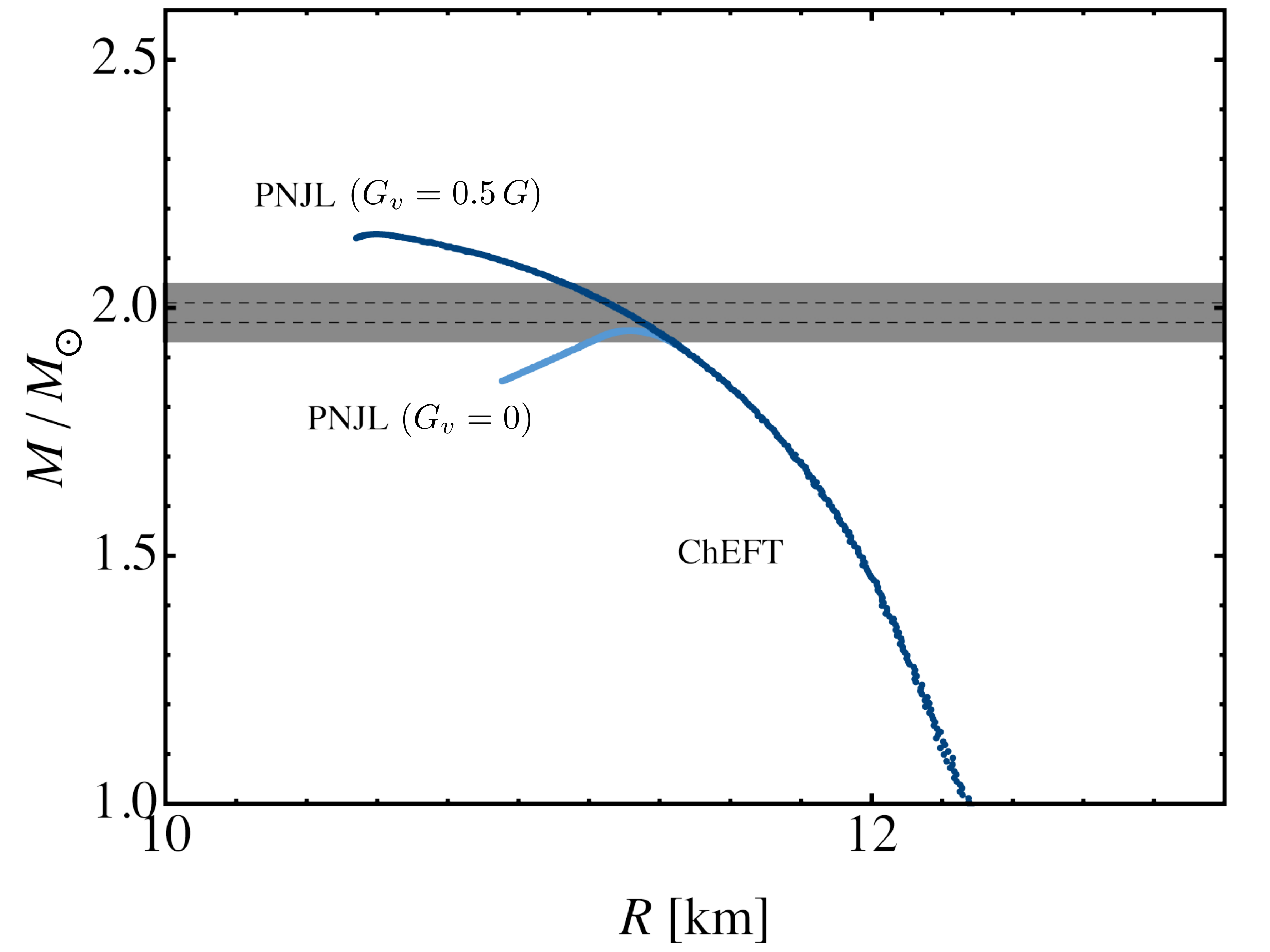}
\end{center}
\vspace{-.5cm}
\caption{Solutions of the TOV equations using the equations of state incorporating a first-order hadron-quark phase transition (see Fig.\,\ref{gibbseosfig}). Mass-radius trajectory lines correspond to different vector coupling strengths $G_v$ as indicated in the figure. The shaded areas are as in Fig.\,\ref{cheftmassradiusfig}. 
}
\label{gibbsmrfig}
\end{figure}

In order to elaborate further on this point, it is instructive to have a look at the density profile of a neutron star with $M=1.95\,M_\odot$, calculated using $G_v=0$ in the hybrid sector. In this case which just barely satisfies the empirical constraints, a possible quark-hadron coexistence domain is restricted to a small part of the inner core within a radius of about 2 km. The central density, $\varrho_c\simeq 5\,\varrho_0$, is only slightly larger than $\varrho_c\simeq 4.8\,\varrho_0$ of the two-solar-mass neutron star reached with the ``conventional" EoS based on chiral EFT.

\begin{figure}[htbp]
\begin{center}
 \includegraphics[width=.45\textwidth]{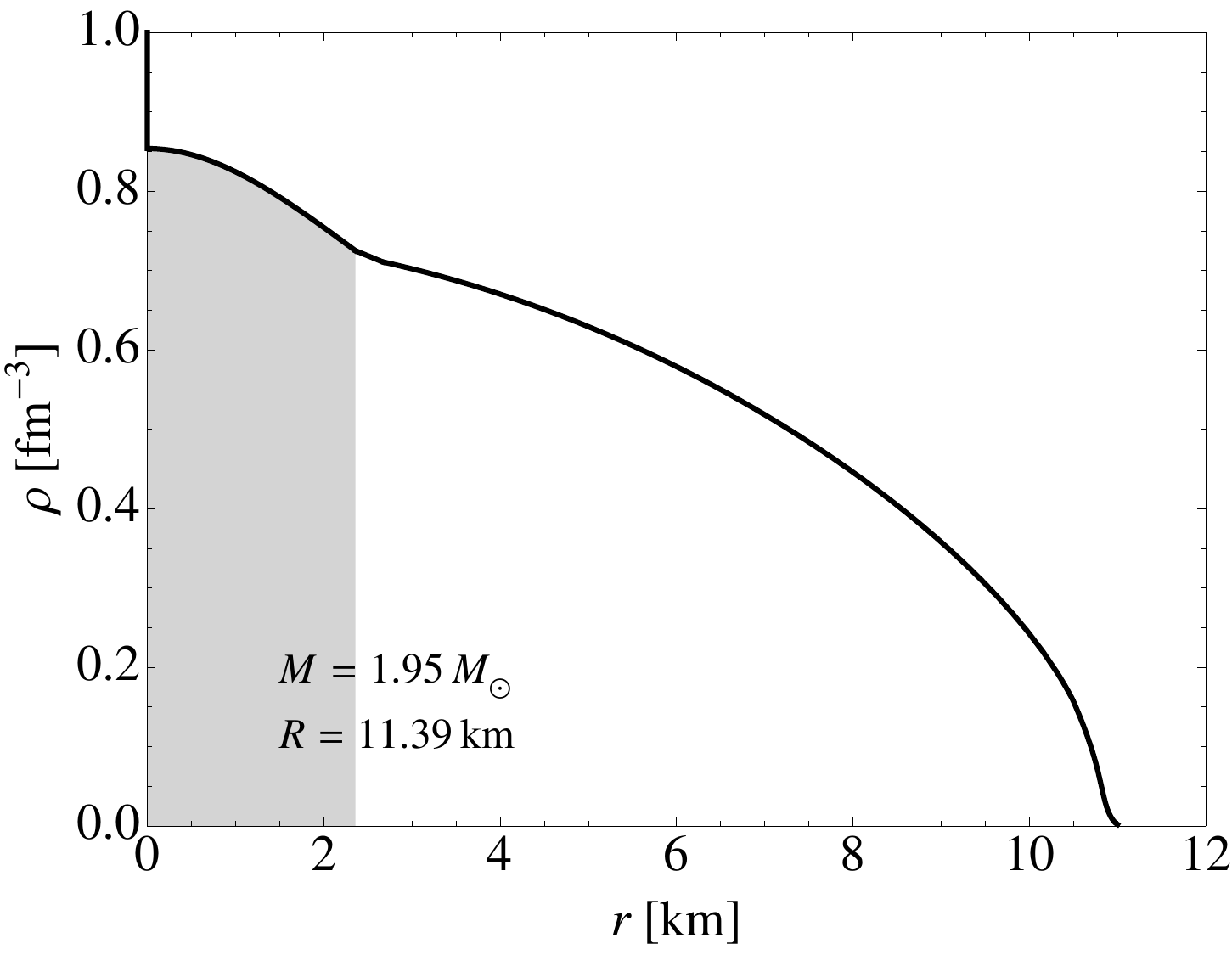}
\end{center}
\vspace{-.5cm}
\caption{Density profile a neutron star with mass $M=1.95\,M_\odot$ and radius $R\simeq 11.4\,{\rm km}$.  The EoS includes the quark-hadron first-order phase transition and no vector interaction, $G_v=0$. The central density is $\varrho_c\approx5\,\varrho_0$. The shaded area shows the onset of the coexistence region, in the inner core within a radius $r \lesssim 2\,{\rm km}$. 
}
\label{gibbsdensityprofilefig}
\end{figure}

We have emphasized repeatedly that the required stiffness of the equation of state keeps the central density of a two-solar-mass neutron star within limits not exceeding typically five times $\varrho_0$. At this point the present model is consistent with the statement in Ref.\,\cite{Lattimer:2010uk}, derived just from causality and the $2 M_\odot$ constraint, that the maximum density cannot exceed $8\,\varrho_0$. The actual bulk baryon densities relevant for most of the material inside a neutron star are significantly lower. Recalling Eq.\,(\ref{nstarmass}) and approximating the energy density roughly as $\epsilon \sim M_N\,\varrho$, one notes
that $r^2\varrho(r)$ rather than the density profile itself matters in the integration of the mass up to the star radius R. For illustration we plot the dimensionless, scaled quantity $(r/R)^2\varrho(r)/\varrho_0$ in Fig.\ref{densityprofile2} and observe that the characteristic bulk densities stay around 2-3$\,\varrho_0$ and hence in a density range where nuclear chiral EFT can well be applied. In this plot the difference between a ``conventional" ChEFT scenario and an EoS including hadron-quark coexistence is almost invisible. A qualitatively similar feature has been noticed in Ref. \cite{Alford:2005}.

\begin{figure}[htbp]
\begin{center}
 \includegraphics[width=.5\textwidth]{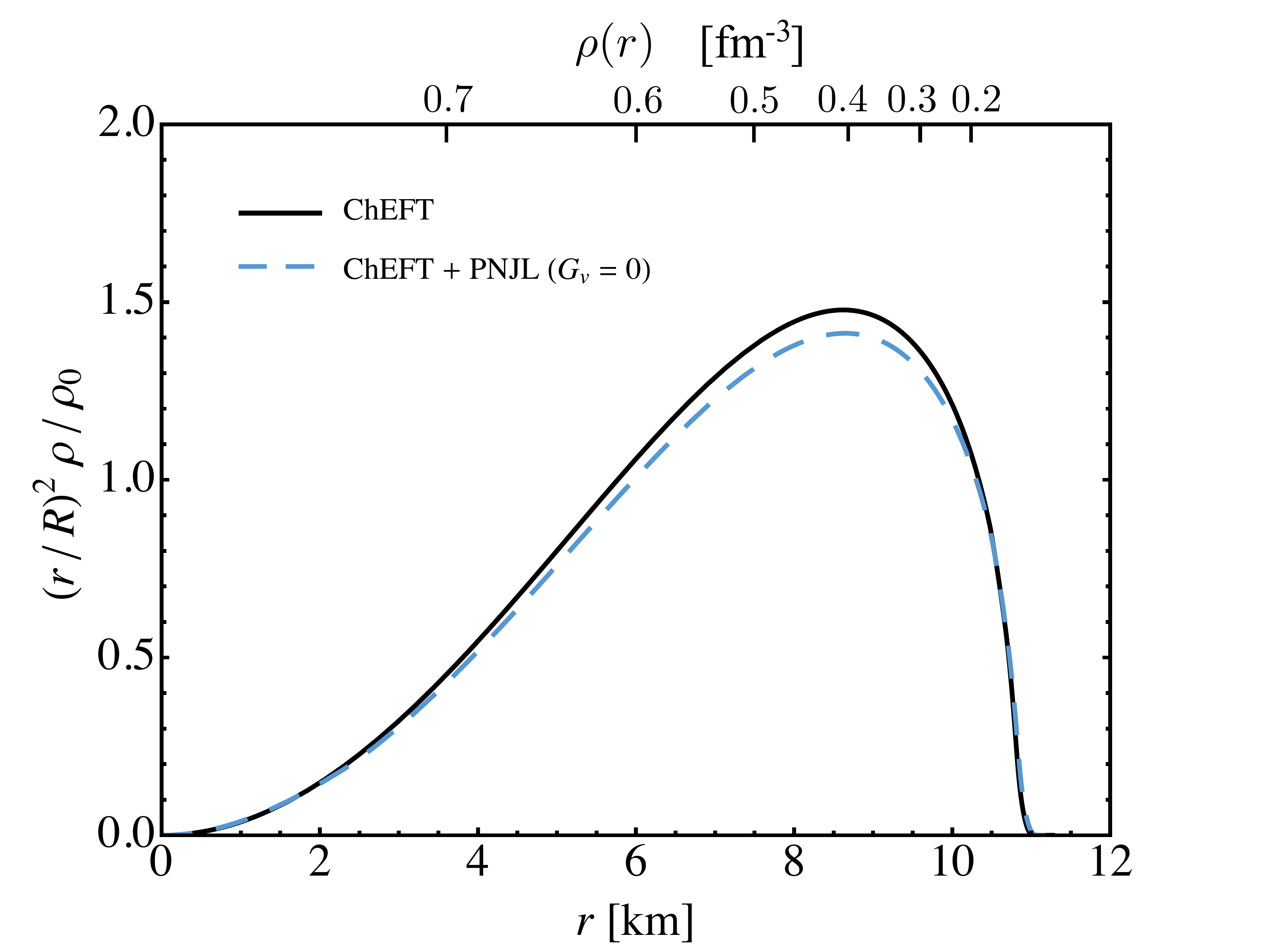}
\end{center}
\vspace{-.5cm}
\caption{Density profiles $\varrho(r)$ multiplied by $r^2$ and scaled with $R^2\varrho_0$ where $R$ is 
the radius of the neutron star and $\varrho_0 = 0.16$ fm$^{-3}$ is the density of normal nuclear matter.
Results are shown for a typical two-solar-mass neutron star. The upper horizontal scale shows the local baryon density. Solid curve: equation of state from nuclear ChEFT ($M = 2\,M_\odot$);
dashed curve: including hadron-quark coexistence in the center of the star ($M = 1.95\,M_\odot$, calculated  using $G_v = 0$).}
\label{densityprofile2}
\end{figure}

The possibility of hadron-quark coexistence has also been studied in Ref.\,\cite{ORWC:2013} using a model that combines a relativistic mean field (RMF) equation of state for the hadronic sector with a non-local PNJL model for quark matter. While the non-local effective interaction between quarks does not make much of a difference compared to the local couplings used in the present work, it should be noted that RMF-based equations of state usually fail to satisfy at least one of the EoS criteria, namely the
requirement of consistency with the most advanced many-body calculations of neutron matter \cite{KTHS:2013,RMP:2014}.

\end{subsubsection}

\end{subsection}

\begin{subsection}{Comments on hyperon admixtures to the EoS}

Admixtures of $\Lambda$ and $\Sigma$ hyperons to the EoS of dense baryonic matter 
in neutron stars have been under discussion for a long time. While $\Sigma$ hyperons 
are not likely to appear since the absence of $\Sigma$ hypernuclei suggests
a weakly repulsive $\Sigma N$ interaction, the low-energy $\Lambda$-nuclear interaction
is attractive. From hypernuclear phenomenology it is known that the $\Lambda$-nuclear
mean field is about half as strong as the Hartree-Fock potential experienced by a nucleon 
in the nuclear medium. 

In neutron star matter, $\Lambda$ hyperons can take over the role of the neutrons when this
becomes energetically favourable at baryon densities exceeding 2-3 times $\varrho_0$. Examples of calculations including hyperons in the EoS can be found in Refs.\,\cite{DSW:2010, WCS:2012}. From these and similar calculations it is now widely accepted that the softening of the equation of state produced by $\Lambda$ admixtures, in the absence of additional repulsive interactions, reduces the maximum mass of a neutron star to values way below two solar masses. 
Additional repulsive forces acting on the hyperons in dense matter are required in order to maintain a sufficiently steep slope of  the pressure $P(\epsilon)$ at high densities.

Our present work features an equation of state for the hadronic sector based on in-medium chiral SU(2)
effective field theory. A fully consistent chiral SU(3) approach to baryonic matter, including both $\Lambda$ and $\Sigma$ hyperons and the complete pseudoscalar meson octet in coupled channels beyond leading order, is not yet available. However, we can present a rough estimate of the admixture of $\Lambda$ hyperon admixtures to the previously derived EoS that combines chiral EFT in the hadronic sector with the three-flavor NJL model for quark matter (see Figs.\,\ref{gibbseosfig},  \ref{gibbsdensityfig}), by simply adding a $\Lambda$ contribution to the energy density, using an attractive mean-field (Hartree) potential adjusted to reproduce hypernuclear data. 

\begin{figure}[htbp]
\begin{center}
 \includegraphics[width=.45\textwidth]{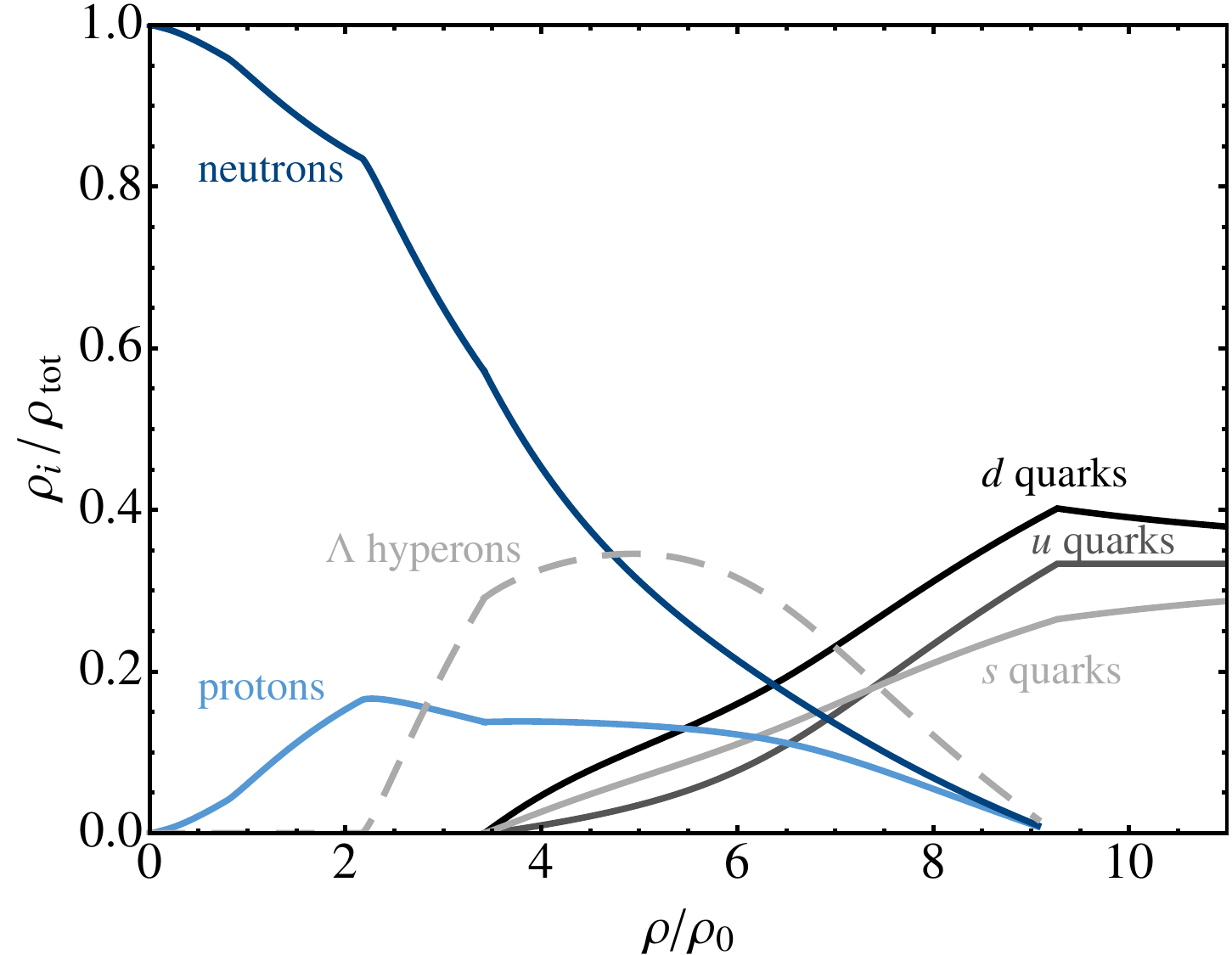}
\end{center}
\vspace{-.5cm}
\caption{Particle ratios as a function of baryon density $\varrho$ (in units of $\varrho_0 = 0.16$ fm$^{-3}$) for the particles indicated, as in 
Fig.\,\ref{gibbsdensityfig} 
but with inclusion of $\Lambda$ hyperons.} 
\label{densityhyperon}
\end{figure}

The result, Fig.\,\ref{densityhyperon}, can be considered as typical and representative for a large class of similar model calculations. The onset of hadron-quark coexistence at $\varrho \simeq 3.5\,\varrho_0$ takes place for a system in which a substantial fraction of neutrons is now substituted by $\Lambda$ hyperons
(implemented here according to the RMF treatment of Ref.\,\cite{GM:1991}). However, the corresponding EoS has now become too soft. It does not satisfy the pertinent constraints and fails to support a two-solar-mass neutron star. For the example shown the maximum neutron star mass is $M_{\mathrm{max}} \simeq 1.5\,M_\odot$. Once $\Lambda$ hyperons are present, the only possiblity to preserve stability of the star within the ``allowed" regions of Fig.\,\ref{constraintseos} appears to be through extra repulsive interactions of the hyperons with the surrounding baryonic medium.

Understanding the origin of such repulsive hyperon-nuclear interactions at high baryon densities
is thus a key issue for the near future.  Advanced Monte Carlo calculations of hyper-nuclear matter \cite{LGP:2013,LPG:2014,LLGP:2014} have recently focused on the role of repulsive three-body $\Lambda NN$ forces. These computations use semi-phenomenological $\Lambda N$ interactions fitted to the available two-body scattering data together with parametrized $\Lambda NN$ potentials constrained by the systematics of $\Lambda$ separation energies in a series of hypernuclei. A sufficiently large $\Lambda NN$ coupling strength in hyper-neutron matter 
\cite{LLGP:2014} does indeed meet the requirement of producing a stiff equation-of-state such that it can satisfy the two-solar-mass constraint.

Steps forward are now taken towards a more systematic foundation of hyperon-nucleon interactions and related three-body forces. An example is the hyperon-nucleon potential in momentum space generated from chiral SU(3) EFT at next-to-leading order (NLO) \cite{HPKMNW:2013}. At this order all two-pion exchange processes are explicitly constructed. Also included is the second-order pion exchange mechanism that drives $\Lambda N \leftrightarrow \Sigma N$ coupled-channels dynamics. This mechanism primarily generates the attractive
mean field that binds the $\Lambda$ in hypernuclei. It is accompanied by smaller repulsive corrections from kaon-exchange Fock terms and from Pauli blocking of the propagating nucleon in the intermediate $\Sigma N$ state of the two-pion exchange process \cite{KW:2005}. The Pauli effect just mentioned acts like an equivalent three-body piece in a description without explicit $\Sigma$,  translating the in-medium $\Lambda N \leftrightarrow \Sigma N$ coupled-channels into effective $\Lambda N$ and $\Lambda NN$ potentials. Such interactions are beginning to be adopted in many-body calculations of hypernuclei \cite{Wirth:2014}. An additional important feature of the chiral SU(3) approach at NLO is the emergence of momentum-dependent repulsive terms \cite{HPKMNW:2013} that grow rapidly with increasing $\Lambda N$ relative momentum. While these terms play only a limited role in $\Lambda$ hypernuclei, they are expected to become increasingly important at the higher baryon densities and Fermi momenta encountered in the center of a neutron star. 

\end{subsection}

\end{section}

\begin{section}{Summary and conclusions}\label{conclusionssection}

The present work contributes to the discussion of the equation of state for dense baryonic matter in view of the by now well established existence of two-solar-mass neutron stars. This study consists of two parts with the following aims: first, to update  the constraints for the pressure as a function of energy density from the new mass determinations together with (less accurate) limits on neutron star radii; secondly, to construct equations of state that are compatible with these observational constraints, while at the same time satisfying the conditions provided by nuclear physics and known properties of nuclear and neutron matter.
 
1. Concerning the first part, the observational constraints determine a band of acceptable neutron star equations of state that are characterized by their pronounced stiffness: at baryon densities $\varrho \simeq 0.8$ fm$^{-3}$, about five times the density of normal nuclear matter in equilibrium, the pressure must at least be $P \gtrsim 150$ MeV fm$^{-3}$ in order to support $2 M_\odot$ neutron stars. This conclusion does not depend on the detailed composition of the matter forming the core of the star. Our results at this point are compatible with related studies reported in 
Refs.\,\cite{Steiner:2010fz,Steiner:2012xt,Lattimer:2013hma,Hebeler:2010jx,Hebeler:2013nza}. 

2. Within the present model investigation of mass-radius trajectories, the stiffness condition on the equation of state has an important implication: the maximum density in the center of the neutron star does not exceed about five times nuclear matter density, corresponding to neutron Fermi momenta less than 0.6 GeV and average kinetic energies of less than 100 MeV. 

3. The modeling of the equation of state in the second part of this work has been performed according to the following criteria. The theory used to construct this equation of state should accurately reproduce:

a) nuclear phenomenology and the thermodynamics of symmetric nuclear matter; 

b) advanced many-body calculations, such as recent Monte Carlo computations, of pure neutron matter; 

c) the symmetry breaking pattern of low-energy QCD and its implications for the nuclear many-body problem. 

In-medium chiral effective field theory is a systematic framework that satisfies these three criteria. The energy density and pressure resulting from this approach at three-loop order does generate the required stiffness of the neutron star equation of state, based on the explicit treatment of two-pion exchange processes, three-body forces, and their in-medium behaviour with proper inclusion of Pauli principle effects.  At its present level of development, in-medium ChEFT is expected to work quite reliably up to about twice to three times the density of normal nuclear matter. Limitations are primarily related to still existent uncertainties in three-body interactions. They amount to errors in the energy per particle of about $5 \%$ at $\varrho_n \sim 2 \,\varrho_0$ and about $20 \%$ at $\varrho_n \sim 3\,\varrho_0$. Further open issues include the role of four-body correlations as they are encountered in the hierachy of chiral effective interactions at higher order. 

4. Nonetheless, ChEFT calculations at three-loop order in the energy density turn out to be consistent with recent Monte Carlo computations of pure neutron matter even up to about four times $\varrho_0$. At the same time, the pertinent baryon densities reached in neutron stars, given the stiffness condition on the EoS, are not extremely high. As pointed out, the bulk material of the star rests primarily on radial regions where the density does not exceed about 2-3 times $\varrho_0$. The physics at such densities is considered to be well accessible to ChEFT methods.

5. Possible scenarios for the appearance of hybrid hadron-quark matter in the deep interior of neutron stars have also been explored in the present work, combining the ChEFT equation of state in the hadronic phase with either continuous or first-order transitions to quark matter. The quark matter component is described schematically in terms of a three-flavor (P)NJL model. Hybrid stars built with such a model are found unable to pass beyond 
the two-solar-mass line unless an additional repulsive vector-current interaction between quarks is introduced in order to generate a sufficiently stiff equation of state. At the same time, such strong repulsion in the quark sector eliminates the first-order chiral phase transition that is characteristic of more basic versions of the NJL model, in favor of a smooth chiral crossover at low temperatures and high densities.

6. The resulting hybrid equation of state, compatible with the criteria under the previous points 1-3, does not feature an extended region of quark matter in the inner core of a neutron star. Likewise, the admixture of strangeness (in the form of hyperons or deconfined strange quarks) is not substantial in such a constrained scenario. The presence of $\Lambda$ hyperons would have again to be accompanied by strongly repulsive $\Lambda N$ and/or $\Lambda NN$ correlations in order to sustain the necessary pressure. 

In summary, the present work supports the idea that neutron stars are indeed predominantly composed of neutrons rather than more exotic forms of matter.

\end{section}

\begin{acknowledgments}

This work has been partially supported by BMBF, by the DFG Cluster of Excellence ``Origin and Structure of the Universe'', and by DFG / NSFC through the Sino-German CRC 110 ``Symmetries and the Emergence of Structure in QCD". We thank Norbert Kaiser, Abishek Mukherjee and Sebastian Schulte{\ss} for many helpful and stimulating discussions. One of us (W. W.) gratefully acknowledges discussions and hospitality during a visit to ITP-CAS in Beijing.

\end{acknowledgments}

\begin{subsection}{APPENDIX: Some details of the PNJL model}

The PNJL grand-canonical potential $\varOmega=-\ln\mathcal{Z}$, with $\mathcal{Z}$ derived from the action $\mathcal{S}_\text{PNJL}$ of Eq.~\eqref{spnjl} in mean-field approximation, is:
	\begin{equation}\label{omegapnjl}
		\begin{aligned}
		\varOmega_\text{MF}&=-\ln\mathcal{Z}_\text{MF}=\dfrac{1}{\beta V}\mathcal{S}_{\text{PNJL, MF}}\\				
	&=-2 T\sum_{a}\sum_{i=u,d,s}\,\sum_{n\in\mathbb{Z}}\int_\varLambda\dfrac{\diff^3 p}{(2\pi)^3}\\&\quad\times\ln\left[(\omega_n^{a,i}+\imu\,\bar v)^2+\vec{p}\,^2+M_i^2\right]\\
			&\quad+\dfrac{\bar\sigma_u^2+\bar\sigma_d^2+\bar\sigma_s^2}{4 G}-\dfrac{\bar v^2}{2 G_V}+\dfrac{K}{2 G^3}\,\bar\sigma_u\,\bar\sigma_d\,\bar\sigma_s\\
		&\quad+\mathcal{U}(\Phi,\bar\Phi;T)\,.
	\end{aligned}
	\end{equation}
This result is found by standard bosonization of Eq.~\eqref{spnjl}, introducing expectation values of the scalar fields, $\bar \sigma_i = -G\,\langle\bar{q}_i q_i\rangle$ ($i\in\{u,d,s\}$), and of the vector field, $\bar v = G_v\,\langle q^\dagger q\rangle$. The dynamically generated (constituent) quark masses are determined by the gap equations
		\begin{equation}\label{gapeqs}
			\begin{aligned}
			M_u&=m_u+\bar\sigma_u+\dfrac{K}{2G^2}\,\bar\sigma_d\,\bar\sigma_s\,,\\
			M_d&=m_d+\bar\sigma_d+\dfrac{K}{2 G^2}\,\bar\sigma_u\,\bar\sigma_s\,,\\
			M_s&=m_s+\bar\sigma_s+\dfrac{K}{2G^2}\,\bar\sigma_u\,\bar\sigma_d\,.
		\end{aligned}
		\end{equation}
These masses (or, equivalently, the scalar mean fields $\bar\sigma_i$) serve as order parameters for the chiral transition.
The shifted Matsubara frequencies $\omega_n^{a,i}$ with $a\in\{0,\pm\},i\in\{u,d,s\}$ are given by:
\begin{equation}
	\begin{aligned}
	\omega_n^{\pm,i}&=\omega_n-\imu\,\mu_i\pm\dfrac{A_4^3}{2}-\dfrac{A_4^8}{2\sqrt{3}}\,,\\
	\omega_n^{0,i}&=\omega_n-\imu\,\mu_i+\dfrac{A_4^8}{\sqrt{3}}\,,
	\end{aligned}
\end{equation}
where $\omega_n=(2n+1)\pi T$, $n\in\mathbb{Z}$, denote the fermionic Matsubara frequencies and the $\mu_i$ are the chemical potentials for each quark species. The thermodynamic potential is written
	\begin{equation}\label{omegapnjldecomp}	\varOmega_\text{MF}=\varOmega_\varLambda+\varOmega_\text{free}+\varOmega_\text{bos}+\mathcal{U}(\Phi,\bar\Phi;T)\,.
	\end{equation}
$\varOmega_\varLambda$ is the fermionic part of Eq.~\eqref{omegapnjl} with quark momenta cut off at $|\vec{p}\,|=\varLambda$; after performing the Matsubara summation one finds:
	\begin{equation}\label{omegalambda}
		\begin{aligned}
		&\varOmega_\varLambda=-6\sum_{i\in\{u,d,s\}}\int_{|\vec{p}\,|\le\varLambda}\dfrac{\diff^3 p}{(2\pi)^3}E^{(i)}(\vec{p}\,)\\
&\quad
-2T\sum_{i\in\{u,d,s\}}\int_{|\vec{p}\,|\le\varLambda}\dfrac{\diff^3 p}{(2\pi)^3}\times\\
		&\quad\left\{\ln\left[1+3(\Phi+\bar\Phi\,\euler^{-\beta E_{-}^{(i)}(\vec{p}\,)})\euler^{-\beta E_{-}^{(i)}(\vec{p}\,)}+\euler^{-3\beta E_{-}^{(i)}(\vec{p}\,)}\right]\right.\\
		&\quad\left.\ln\left[1+3(\bar\Phi+\Phi\,\euler^{-\beta E_{+}^{(i)}(\vec{p}\,)})\euler^{-\beta E_{+}^{(i)}(\vec{p}\,)}+\euler^{-3\beta E_{+}^{(i)}(\vec{p}\,)}\right]\right\}
		\end{aligned}
	\end{equation}
(with $\beta=1/T$). The quasiparticle energies of the quarks are ($i\in\{u,d,s\}$):
	\begin{equation}
	\begin{aligned}
		E_\pm^{(i)}(\vec{p}\,)=\sqrt{\vec{p}\,^2+M_i^2}\pm (\mu_i-\bar v)\,.
	\end{aligned}
	\end{equation}
The potential $\varOmega_\text{free}$ is the contribution of a gas of quarks with momenta above the cutoff $\Lambda$. These high-momentum quarks have their current-quark masses and do not interact. This added contribution makes sure that recover the correct Stefan-Boltzmann limit is recovered for the pressure and the energy density. 
The last two pieces in Eq.\,(\ref{omegapnjldecomp}) are: 
	\begin{equation}
		\varOmega_\text{bos}=\dfrac{\bar\sigma_u^2+\bar\sigma_d^2+\bar\sigma_s^2}{4 G}-\dfrac{\bar v^2}{2 G_V}+\dfrac{K}{2 G^3}\,\bar\sigma_u\,\bar\sigma_d\,\bar\sigma_s
	\end{equation}
and the Polyakov effective potential \eqref{upolyakov}.

\end{subsection}

\end{document}